\documentclass[prd,a4paper,superscriptaddress,nofootinbib]{revtex4}
\usepackage{amsmath,amssymb,epsfig,natbib,bm}

\begin{document}

%conditional stuff for whether astro-ph or PRD.
%Anything about colours include the alternative with \BW

%\newcommand{\BW}[2]{#1}  %BW for PRD
\newcommand{\BW}[2]{#2}   %Colour

%%%%%%%%%%%%%%%%%%%%%%%%%%%%%%%%%%%%%%%%%%%%%%%%%%%%%%%%%%%%%%%%%%
%                       Own commands                             %
%%%%%%%%%%%%%%%%%%%%%%%%%%%%%%%%%%%%%%%%%%%%%%%%%%%%%%%%%%%%%%%%%%

\newcommand{\Mpc}{\text{Mpc}}
\newcommand{\half}{{\textstyle \frac{1}{2}}}
\newcommand{\third}{{\textstyle \frac{1}{3}}}
\newcommand{\numfrac}[2]{{\textstyle \frac{#1}{#2}}}
\newcommand{\ra}{\rangle}
\newcommand{\la}{\langle}
\renewcommand{\d}{\text{d}}

\newcommand{\cla}{\mathcal{A}}
\newcommand{\clb}{\mathcal{B}}
\newcommand{\clc}{\mathcal{C}}

\newcommand{\cle}{\mathcal{E}}
\newcommand{\clf}{\mathcal{F}}
\newcommand{\clg}{\mathcal{G}}
\newcommand{\clh}{\mathcal{H}}
\newcommand{\cli}{\mathcal{I}}
\newcommand{\clj}{\mathcal{J}}
\newcommand{\clk}{\mathcal{K}}
\newcommand{\cll}{\mathcal{L}}
\newcommand{\clm}{\mathcal{M}}
\newcommand{\cln}{\mathcal{N}}
\newcommand{\clo}{\mathcal{O}}
\newcommand{\clp}{\mathcal{P}}
\newcommand{\clq}{\mathcal{Q}}
\newcommand{\clr}{\mathcal{R}}
\newcommand{\cls}{\mathcal{S}}
\newcommand{\clt}{\mathcal{T}}
\newcommand{\clu}{\mathcal{U}}
\newcommand{\clv}{\mathcal{V}}
\newcommand{\clw}{\mathcal{W}}
\newcommand{\clx}{\mathcal{X}}
\newcommand{\cly}{\mathcal{Y}}
\newcommand{\clz}{\mathcal{Z}}
% Miscellaneous
\newcommand{\CMBFAST}{\textsc{cmbfast}}
\newcommand{\CAMB}{\textsc{camb}}
\newcommand{\Omtot}{\Omega_{\mathrm{tot}}}
\newcommand{\Omb}{\Omega_{\mathrm{b}}}
\newcommand{\Omc}{\Omega_{\mathrm{c}}}
\newcommand{\Omm}{\Omega_{\mathrm{m}}}
\newcommand{\omb}{\omega_{\mathrm{b}}}
\newcommand{\omc}{\omega_{\mathrm{c}}}
\newcommand{\omm}{\omega_{\mathrm{m}}}
\newcommand{\Omdm}{\Omega_{\mathrm{DM}}}
\newcommand{\Omnu}{\Omega_{\nu}}

\newcommand{\Oml}{\Omega_\Lambda}
\newcommand{\OmK}{\Omega_K}

\newcommand{\Hunit}{~\text{km}~\text{s}^{-1} \Mpc^{-1}}
\newcommand{\Gyr}{{\rm Gyr}}

\newcommand{\nrun}{n_{\text{run}}}

\newcommand{\lmax}{l_{\text{max}}}

\newcommand{\zre}{z_{\text{re}}}
\newcommand{\mpl}{m_{\text{Pl}}}

\newcommand{\vv}{\mathbf{v}}
\newcommand{\vd}{\mathbf{d}}
\newcommand{\mN}{\bm{N}}
\newcommand{\eV}{\,\text{eV}}
\newcommand{\vtheta}{\bm{\theta}}
%%%%%%%%%%%%%%%%%%%%%%%%%%%%%%%%%%%%%%%%%%%%%%%%%%%%%%%%%%%%%%%%%%
%                       Title matter                             %
%%%%%%%%%%%%%%%%%%%%%%%%%%%%%%%%%%%%%%%%%%%%%%%%%%%%%%%%%%%%%%%%%%

% title and affiliations
\title{Cosmological parameters from CMB and other data: a Monte-Carlo approach}

\author{Antony Lewis}
 \email{Antony@AntonyLewis.com}
 \affiliation{DAMTP, CMS, Wilberforce Road, Cambridge CB3 0WA, UK.}

\author{Sarah Bridle}
 \email{sarah@ast.cam.ac.uk}
 \affiliation{Institute of Astronomy, Madingley Road, Cambridge, CB3 0HA, UK.}

\begin{abstract}
\vspace{\baselineskip}
We present a fast Markov Chain Monte-Carlo exploration of cosmological
parameter space. We perform a joint analysis of results from recent CMB 
experiments
and provide parameter constraints, including $\sigma_8$, from the CMB
independent of other data.
We next combine data from the CMB, HST Key Project, 2dF galaxy redshift survey,
 supernovae Ia and big-bang nucleosynthesis.
The Monte Carlo method allows the rapid investigation of
a large number of parameters, and we present results from 6 and 9 parameter
analyses of flat models, and an 11 parameter analysis of non-flat models.
Our results include constraints on the neutrino mass ($m_\nu \alt 0.3 \eV$), 
equation of state of the
dark energy, and the tensor amplitude, as well as demonstrating the
effect of additional parameters on the base parameter constraints.
In a series of appendices we describe the many uses of importance 
sampling, including computing results from new data and 
accuracy correction of results generated from an approximate method.
We also discuss the different ways of converting
parameter samples to parameter constraints, the effect of the prior,
assess the goodness of fit and consistency, and describe the use of analytic marginalization over
normalization parameters.
\end{abstract}

%Cosmology, Obs cosmology, Monte-Carlo methods,neutrino mass and
%mixing

\maketitle

%\keywords{cosmic microwave background --- cosmology: theory}

%%%%%%%%%%%%%%%%%%%%%%%%%%%%%%%%%%%%%%%%%%%%%%%%%%%%%%%%%%%%%%%%%%
%                         Main body                              %
%%%%%%%%%%%%%%%%%%%%%%%%%%%%%%%%%%%%%%%%%%%%%%%%%%%%%%%%%%%%%%%%%%

\section{Introduction}

There is now a wealth of data from cosmic microwave background (CMB)
observations and growing amount of information on large scale
structure from a wide range of sources.
We would like to extract the maximum amount of information 
from this data, usually
conveniently summarized by estimates of a set of cosmological
parameter values. With high quality data one can constrain a large number of
parameters, 
which in principle allows us not
only to put estimates and error bars on various quantitative parameters,
but also to address more fundamental qualitative questions: Do we
observe primordial gravitational waves? Do the neutrinos have a
cosmologically significant mass? Is the universe flat? Are the standard 
model parameters
those that best account for the data? 
In addition we can also assess the consistency of the different data 
sets with respect to a cosmological model.

Recent work involving parameter estimation from the CMB includes 
Refs.~\cite{Lahav02,Efstathiou02,KnoxCS01,Wang01,Netterfield01,Bernardis01,Pryke02,Sievers02,Percival02,Tegmark02}.
In this paper
we employ Markov Chain Monte Carlo (MCMC) techniques~\cite{MCMC97,MacKayBook,Neil93}, as advocated for
Bayesian CMB analysis in Ref.~\cite{Christensen01}
and demonstrated by Refs.~\cite{KnoxCS01,VSA4}.
By generating a
set of MCMC chains we can
obtain a set of independent samples from the posterior distribution of the
parameters given the data. From a relatively small number of
these samples one can estimate the marginalized posterior
distributions of the parameters, as well as various other statistical
quantities. The great advantage of the MCMC method is that it scales,
at its best, approximately linearly 
with the number of parameters, allowing us to
include many parameters for only small additional computational
cost. The samples also probe the shape of the full posterior, giving
far more information that just the marginalized distributions.

Throughout we assume models with purely adiabatic Gaussian primordial
perturbations in the growing mode (our approach could easily be generalized to
include isocurvature modes), three neutrino species,
non-interacting cold dark matter, and standard general relativity.
We compute all theoretical (CMB and matter power spectrum) 
predictions numerically using using the fast Boltzmann code \CAMB~\cite{Lewis99} (a
parallelized version of \CMBFAST~\cite{Seljak96}). Our results are
therefore limited by the accuracy of the data and 
could be generalized very easily to include any
additional parameters that can be accounted for by modification of a Boltzmann code.

In section~\ref{CMBonly} we present constraints from the latest CMB data,
illustrating the MCMC method. We defer
a brief introduction to the MCMC method and
a description of our implementation and terminology to
Appendix~\ref{MCMC}. The method of importance sampling is 
also illustrated in section~\ref{CMBonly}, and is
described in detail
Appendix~\ref{importance}, where we explain how it can be used to 
take into account
different priors on the parameters, new data, and for accurate but fast
estimation given a good approximation to the theoretical model
predictions. 
We add large scale structure, supernova and nucleosynthesis constraints
in section~\ref{alldata} so that more parameters can be constrained.
We compare flat models with `inflationary' priors 
(9 cosmological parameters) and then a more general models (11
cosmological parameters). 
This includes constraints on the neutrino mass, equation of state of the
dark energy, and the tensor amplitude. In addition our results show
the sensitivity of constraints on the standard parameters to variations
in the underlying model.

\section{CMB constraints}
\label{CMBonly}

%%%%%%%%%%%%%%%%%%%%%%%%%%%%%%%%%%
\begin{figure}
\begin{center}
\epsfig{figure=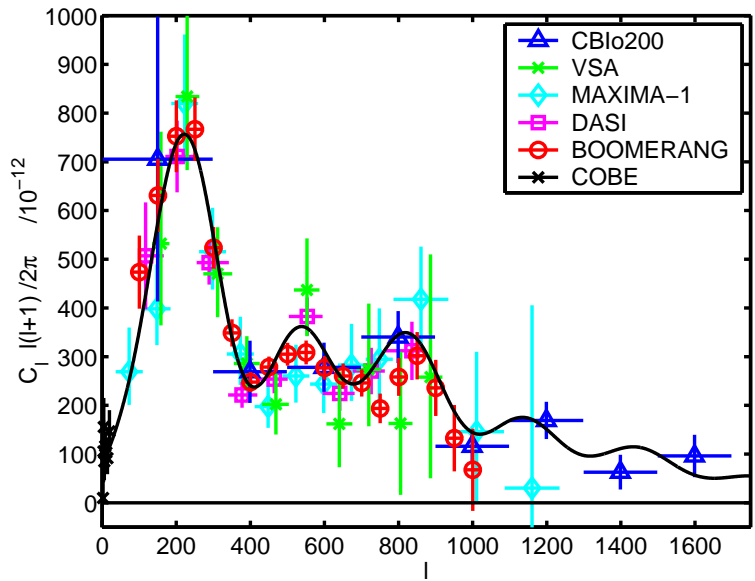,angle=0,width=10cm}
\caption{
The CMB temperature anisotropy band-power data used in this paper.
The line shows the model with the parameters at their mean values,
given all data
after marginalizing in 6 dimensions 
(i.e. first column of Table.~\ref{inflationtable}).
\label{cmbdata}
}
\end{center}
\end{figure}
%%%%%%%%%%%%%%%%%%%%%%%%%%%%%%%%%%

We use the results of the COBE~\cite{COBE},
BOOMERANG~\cite{Netterfield01}, MAXIMA~\cite{Hanany00},
DASI~\cite{Halverson01}, VSA~\cite{VSA3} and CBI~\cite{Pearson02,CBIdata} observations in the form of band power
estimates for the temperature CMB power spectrum. 
These data are plotted in Fig. \ref{cmbdata}. 
The very small angular scale results from CBI have been discussed
extensively in~\cite{Pearson02,Bond02} and do not fit the linear predictions
of standard acoustic oscillation models.
Therefore for the purposes of this paper we assume that the small
scale power observed by CBI has its origin in non-linear or
non-standard small scale effects that we do not attempt to model, 
and so use only the mosaic (rather than deep pointing) data points throughout.
In addition we use only the first 8 points
 ($\ell\alt2000$) from the odd
binning of the mosaic fields since above that the noise in the band powers becomes
much larger than the prediction for the class of models we consider.

For COBE we use the offset-lognormal band-powers and covariance
matrix from RADPACK~\cite{Bond98}.
For DASI, VSA and CBI we also use the offset-lognormal band powers and integrate
numerically over an assumed Gaussian calibration uncertainty.
For BOOMERANG, MAXIMA we assume top hat window functions and uncorrelated 
Gaussian band-power likelihood distributions, and
marginalize analytically over the calibration and beam uncertainties 
assuming they are also Gaussian~\cite{Bridle01}. 
We assume a correlated calibration uncertainty of $10\%$ on the CBI and
VSA data (neglecting the $\sim 3\%$ uncorrelated difference in
calibration), but otherwise assume all the observations are independent.
Using the sets of samples obtained it is a simple matter to account for small
corrections when the required information is publicly available (see
Appendix~\ref{importance}).

The base set of cosmological parameters we sample over 
are $\omb = \Omb h^2$ and $\omc \equiv \Omc h^2$, the
physical baryon and CDM densities relative to the critical density,
$h = H_0/(100 \Hunit)$, the Hubble parameter, 
$\OmK \equiv 1 - \Omtot$ measuring the spatial curvature, $\zre$,
the redshift at which the reionization fraction is a half
\footnote{The CMB temperature anisotropy is very insensitive to the
duration of reionization epoch, and we also neglect the small effect
of Helium reionization and inhomogeneities.},
$A_s$,
measuring the initial power spectrum amplitude
and $n_s$, the spectral index of the initial power spectrum.
We derive $\Oml$, the ratio of the critical density in the form of
dark energy, from the constraint $\OmK + \Oml + \Omm = 1$
(where $\Omm \equiv \Omc + \Omb$ is the total matter density in units of
the critical density).
Throughout
we use at least the priors that $4 < \zre < 20$, 
$0.4< h <1.0$, $-0.3 < \OmK < 0.3$, 
$\Oml>0$, and that the age of the universe, $t_0$, is
$10 \,\Gyr < t_0 < 20 \,\Gyr$.
The significance of this base set is that this defines the
Bayesian priors:
there is a flat prior on each parameter of the base set.
We discuss later how we assess the significance of these priors,
and highlight our main results, which are largely independent of the
priors.
(We choose $h$ as a base
parameter since the HST Key Project provides a direct constraint on
this quantity, whereas there are no direct constraints on,
e.g. $\Oml$; see Appendix~\ref{constraints} for discussion).
The above additional constraints on $h$, $\Oml$, $\OmK$ and the age
have little effect on the joint results since the cut-off values are well into
the tails of the distribution. However for the purpose of the
Monte-Carlo it is very convenient to be able to quickly reject models in the
extreme tails without having to compute the theoretical predictions. 
 
\subsection*{MCMC illustration}

%%%%%%%%%%%%%%%%%%%%%%%%%%%%%%%%%%
\begin{figure}
\begin{center}
\begin{tabular}{cc}
\BW{
\epsfig{figure=samples_bw.ps,angle=0,width=7.4cm}
}{
 \epsfig{figure=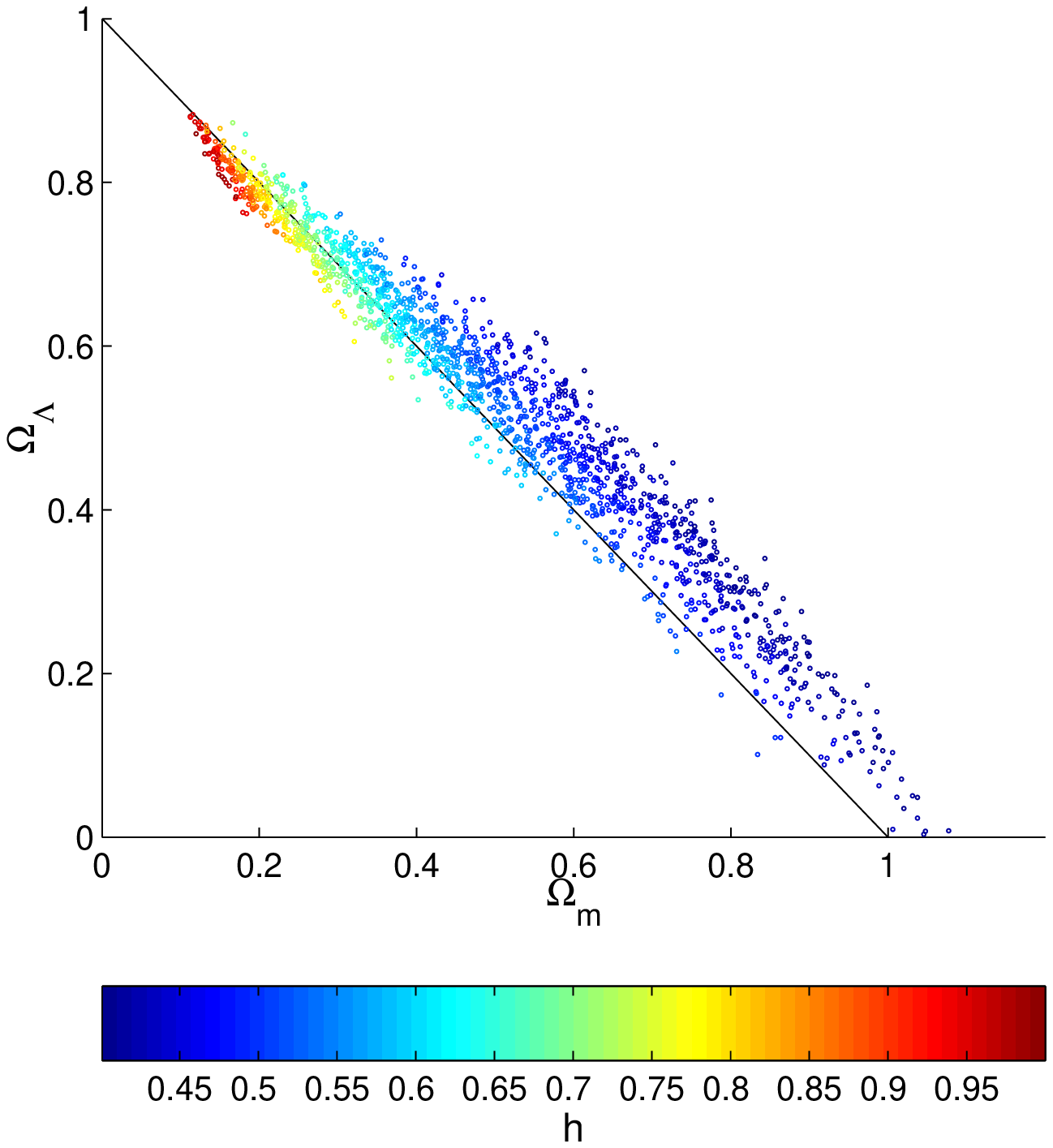,angle=0,width=7.4cm} 
}
\epsfig{figure=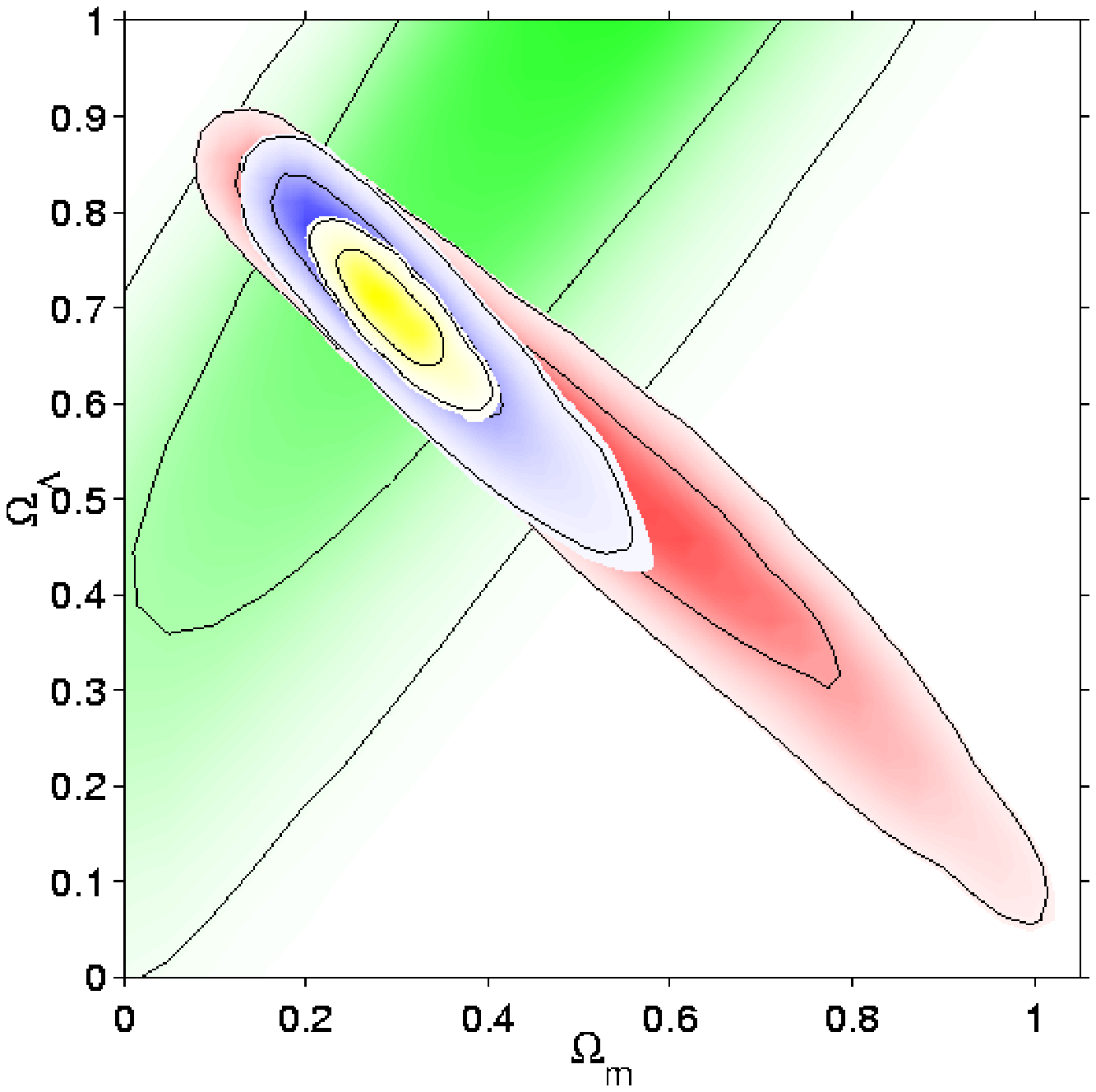,angle=0,width=8cm} 
\end{tabular}
\caption{
\label{cmbonly}
\emph{Left:} 2000 samples from the posterior distribution of the
parameters plotted by their $\Omm$ and $\Oml$ values.
Points are \BW{shaded}{colored} according to the value of $h$ of each sample,
and the solid line shows the flat universe parameters.
We assume the base parameter set with broad top-hat priors. \\
\emph{Right:}
bottom layer\BW{}{ (green)}: supernova constraints;
next layer up\BW{}{ (red)}: CMB data alone;
next\BW{}{ (blue)}: CMB data plus HST Key Project prior;
top layer\BW{}{ (yellow)}: all data combined (see text).
68 and 95 per cent confidence limits are shown.
}
\end{center}
\end{figure}
%%%%%%%%%%%%%%%%%%%%%%%%%%%%%%%%%%

An MCMC sampler provides an efficient way to generate a list of
samples from a probability distribution (see Appendix~\ref{MCMC} for an explanation). All that
is required is a function for calculating the probability given a set
of parameter values.
A single sample is a coordinate in the n-D parameter space, and the sampling
method ensures that the number density of samples is asymptotically
proportional to the probability density.
As an illustration, in the left hand panel of Figure~\ref{cmbonly} we show 
the values of $\Omm=\Omb+\Omc$, and
$\Oml$, for samples collected from an
MCMC run using the CMB data and base parameters discussed in the
previous paragraphs.

Note that although $\Omm$ is not one of our base set of parameters,
it is simple to find probabilities as a function of $\Omm$ 
by taking the parameter values of each sample and deriving the
corresponding values of $\Omm$.
Since the MCMC method produces samples from the full posterior,
it follows that
 the number density of samples
in this two-dimensional plane is proportional to the 
probability density of the two parameters
(marginalized 
over all the other parameters; note that this refers to the fully marginalized probability density
rather than any conditional or projected probability density).
The familiar direction of CMB degeneracy 
along the flat universe line is apparent. 
The \BW{darkness}{colors} of the dots indicate the Hubble constant value for 
each sample, as given in the \BW{gradient}{color} bar. 
This shows that the high $\Omm$ tail of samples are due 
entirely to low $h$ regions of parameter space, illustrating
the point made in e.g.~\cite{VSA4} 
that a Hubble constant prior
alone combined with the CMB can put useful limits on $\Omm$ and
$\Oml$ without the need for supernova constraints.

The likelihood as a function of position in the $\Omm$-$\Oml$ plane
from CMB data alone is shown by the \BW{broad}{red} contours in the right hand
panel of Figure~\ref{cmbonly}. 
Note that compared to some 
other CMB only plots the contours close
slightly at high $\Omm$, which is due to our lower limit on the Hubble 
constant of $h>0.4$.
The 2-D likelihoods shown are calculated from the samples
by estimating the probability
density at a grid of points using a Gaussian weighting kernel and
then using interpolation to draw the contour lines. 
This produces a plot very similar to that
obtained using a grid integration calculation of the
marginalized posterior (assuming there are no small scale features in
the posterior).

Extra data can be taken into account quickly by re-weighting the 
samples, a technique known as {\it importance sampling} (described
in detail in Appendix~\ref{importance}). The posterior is simply
re-calculated at each sample position, and a weight assigned to the
sample according to the ratio of the new posterior to the old posterior.
For example, using a Gaussian HST Key Project~\cite{Freedman01} 
prior on the Hubble constant $h = 0.72 \pm 0.08$ we obtain 
the \BW{second set of}{blue} contours 
plotted on the right hand panel of Figure~\ref{cmbonly}.  By using the weighted number of points as a function of $\OmK$
we find the marginalized result $\OmK =0.00 \pm 0.03$, and therefore
that the universe is close to flat (given our assumptions).

Using importance sampling we have checked that the CMB datasets are
consistent using the hyperparameter method, as described in Appendix~\ref{consistency}.

\subsection*{Quantitative constraints from CMB data alone}

Above we illustrated the MCMC method in the $\Omm$-$\Oml$ plane
for a 7 parameter cosmological model. Since there is good observational evidence and theoretical motivation
for flat models, we now fix $\OmK=0$ giving 6 base parameters.

The only parameters
that can be reasonably disentangled are $\omb$ and $n_s$, from
the heights of the 2nd and 3rd acoustic peaks relative to the first.
These constraints are given in the first two rows
of Table~\ref{cmbflatpars}
and are in accordance with previous similar analyses and 
nucleosynthesis constraints~\cite{Burles01}.
However note that 
%Ain fact 
these two parameters remain
significantly correlated, so that 
%Ain fact 
more detailed information is contained in the result
$n_s - (\omb-0.022)/0.043 =
0.98\pm 0.04$, $n_s + (\omb-0.022)/0.043 = 0.98\pm 0.09$.

\begin{table}
\vbox{\center{\centerline{
\begin{tabular}{|c|r@{ $\pm$ }l|r@{ $\pm$ }l|r@{ $\pm$ }l|}
\hline 
Parameter                  & \multicolumn{2}{|c|}{pre VSA/CBI} &
\multicolumn{2}{|c|}{+VSA+CBI } & \multicolumn{2}{|c|}{all data}\\
\hline  
%A Have rounded some exponents and updated values
%for some reason I get H^4.7 in v2 when I work it out now.
$\Omb h^2$    &   $0.0215$ & $0.0022$   & $ 0.0216 $ & $ 0.0022 $ &$0.0212 $ & $ 0.0014$\\
$n_s$             &  $ 0.985$ & $0.051$   & $0.982 $ & $ 0.050 $ &$0.980 $ & $ 0.037$\\ 
$v_1 \equiv \Omm h^{2.4} (\sigma_8 e^{-\tau}/0.7)^{-0.85}$  & $0.114$
& $0.0052 $& $0.113 $ & $ 0.0048$ & $0.111 $ & $ 0.0038$ \\
$v_2 \equiv \sigma_8 e^{-\tau} (h/0.7)^{0.5} (\Omm/0.3)^{-0.08}$ &
$0.71$ & $0.07$ & $0.70 $ & $ 0.06$ & $0.70 $ & $ 0.04$ \\
\hline
\end{tabular}
}}}
\caption{
Marginalized parameter constraints (68 per cent confidence) 
on four well constrained parameters (see text) from 
two combinations of CMB data  assuming a flat universe and
varying [$\Omb h^2$, $\Omc h^2$, $h$, $n_s$, $z_{\rm re}$, $A_s$].
Extra digits are inserted to help comparison between pre and post 
VSA and CBI. 
For comparison the last column gives the results including 
HST, 2dF, BBN and SnIA data. 
}
\label{cmbflatpars}
\end{table}

Qualitatively, having `used up' constraints from the 2nd and
3rd acoustic peak heights to find $\Omb h^2$ and
$n_s$ we might estimate that there remain
three more pieces of information in the CMB data, from the 
large scale (COBE) amplitude, the first peak height and the
first peak position.
We have four remaining parameters
e.g. parameterised
by $\sigma_8$, $\Omm$, $h$ and $\tau$ ($\tau$ is the optical 
depth to reionization\footnote{Assuming rapid reionization the optical depth can be calculated from $\zre$
using $\tau = \sigma_T\int_0^{\zre} \d z n_e(z)/[(1+z)^2 H(z)]$,
where $n_e$ is the number density of electrons and $\sigma_T$ is the
Thompson scattering cross-section.
For flat models with cosmological constant
$\tau \approx 0.048 \,\Omb\, h \,\Omm^{-0.47}\, \zre^{1.42} $ (to a
few percent over the region of interest), though we do
not use this in our analysis.}, 
and $\sigma_8$ is the 
root mean square mass perturbation 
in $8 h^{-1}\Mpc$ spheres today assuming linear evolution).
Since $\Omb h^2$ and $n_s$ are not too correlated with these
parameters, we marginalize over them and explore the 
remaining four dimensional parameter space.

From the set of samples it is straightforward to
perform an independent component analysis to identify the well
determined orthogonal parameter combinations. Calculating the 
 eigenvectors of the correlation matrix 
in the logs of $\sigma_8 e^{-\tau}$, $\Omm$ and $h$
we find the two combinations given in the bottom two rows of 
Table~\ref{cmbflatpars}.
The errors are too large on any third constraint to be
worth quoting.
We expect $\sigma_8$ to be roughly degenerate with $e^{-\tau}$
because the CMB power on scales 
smaller than the horizon size at reionization
($\ell \agt 20$)
is damped by a factor $e^{-2\tau}$, and 
$\sigma_8^2$ scales with the power in the primordial perturbation.
We find that marginalized values of $v_1$ and $v_2$ are independent
of $\tau$ to better than 2 per cent for $0.02<\tau < 0.14$  ($4 < \zre < 16$).
As demonstrated in
Appendix~\ref{constraints} these constraints are almost independent of
the choice of base parameters (which define the prior).

If we marginalize over $\sigma_8$ and $\zre$ we find
the principle component $\Omm h^{2.9} = 0.096\pm 9\%$, consistent
with the constraint found and discussed in Ref.~\cite{Percival02}.
However since this quantity is significantly correlated with the
amplitude we have quoted a tighter constraint ($4\%$)
in Table~\ref{cmbflatpars} 
by including the $\sigma_8 e^{-\tau}$ dependence in our results.\footnote{
Note that the fractional errors depend on the choice of
normalization for the log eigenvectors; here we have chosen to
normalize so the exponent of $\Omm$ is unity.} 

While restricted to this relatively small parameter space we take
this opportunity to investigate the impact
of the new VSA and CBI results. 
Using importance sampling we compare the results with and without the
VSA/CBI data in
Table~\ref{cmbflatpars}.
For simplicity we assume the same power law approximation for 
the combination of $\sigma_8 e^{-\tau}$, $h$ and $\Omm$ as derived above.
The peaks move by a fraction of the error, and the error bars are
fractionally smaller.

\section{Additional cosmological constraints}
\label{alldata}

The CMB data alone can only provide 
a limited number of constraints,
so before extending the parameter space to make
full use of the Monte-Carlo method it is 
useful to include as much relatively reliable 
data as possible. 
Care must be taken, since
some published parameter constraints assume particular values 
for parameters that we wish to vary. As a simple example, the 
Supernova Cosmology Project has made available constraints in
$\Omm$-$\Oml$ space, which could only be used if the 
dark energy is a cosmological constant.
Fortunately the full supernova data are available, which we use (described below). 
However is is not always practical to use the full observed data
and it may be best to simply increase the  error bars 
to encompass systematic effects due to other parameters. This
ensures that the resulting MCMC chains will cover all of the 
relevant parameter space, and can
then be importance sampled later with a tighter constraint.
If the
chains are generated too tightly constrained one cannot recover
information about the excluded parameter space.

Nucleosynthesis constraints suggest $\omb \approx
0.02$~\cite{Burles01}, and we assume the Gaussian prior $\omb = 0.020
\pm 0.002$,  (1$\sigma$) 
which is somewhat broader than the error quoted in
Ref.~\cite{Burles01} to allow for differences with other estimations. We include Type 1A supernovae data
from~\cite{Perlmutter98}, using the effective magnitudes and errors from the 54
supernovae that they included in the primary 
fit (fit C). We marginalize
analytically with a flat prior on the intrinsic magnitudes, which is
equivalent 
to evaluating the likelihood
at the best fit value (see Appendix~\ref{marge}). We neglect the small
correlations but multiply the log likelihood by $50/54$ to account for
the effective degrees of freedom quoted in Ref.~\cite{Perlmutter98}.
We use the galaxy power spectrum from the first 147,000 redshifts of
the 2dF galaxy redshift survey, using 
the scales 
$0.02 <  k/(h \Mpc^{-1}) < 0.15$ where non-linear effects were found to be negligible~\cite{Percival02}.
We assume that this is directly proportional to the matter power spectrum
at $z=0$, in other words that the bias and evolution are scale
independent 
and also that the redshift space distorted power spectrum is
proportional to the real space power spectrum
(on the large scales used). 
We assume a flat prior on the proportionality constant
and marginalize analytically as described in Appendix~\ref{marge}.
We also use the HST Key Project prior on the Hubble constant as
discussed earlier.

The top contours in the
right hand panel of Figure~\ref{cmbonly} shows the effect of the
full set of constraints on the basic 7 parameter model, with the
combined constraint on the curvature becoming $\OmK = 0.00 \pm 0.02$.  
For the 6 parameter flat models the extra data constrains most of the
parameters rather well.
Table~\ref{cmbflatpars} shows that the new constraints
are very consistent with those from the CMB alone. 
The $v_2$ constraint on $\sigma_8$ of Table~\ref{cmbflatpars} is 
slightly changed
and becomes  $\sigma_8 e^{-\tau}(h/0.67)^{0.58} = 0.72 \pm 0.05$ 
almost independently of $\Omm$.  
The marginalized results on all parameters are
shown in Table~\ref{inflationtable}.
The Hubble parameter is shifted to slightly lower values relative to
the HST Key Project constraint we used.
The matter and dark energy densities are spot on the
popular $0.3, 0.7$ model, although with error bars of $0.05$
at 68 per cent confidence.
The combination $\Omm h$ is slightly tighter but
has the same central value 
as quoted in Ref.~\cite{Percival01}
using 2dF data alone and assuming $n_s=1$.

The 
$\sigma_8$ result depends 
on the range used for the reionization redshift $\zre$ since
the data used mostly constrains
the combination $\sigma_8 e^{-\tau}$
rather than $\sigma_8$ on its own. 
Our prior of $4 <\zre <20$ should
be reasonably conservative, however we also quote the
result for $\sigma_8 e^{-\tau}$.
Looking at the maximum and minimum values contained in the 95 percent
confidence region of the full n-dimensional space (see Appendix C)
we find $0.62 < \sigma_8 \exp(0.04-\tau) < 0.92$.
This may be compared to values of $\sigma_8$ found by other
independent methods and could in principle be combined with these
methods to estimate the optical depth (e.g. see 
Ref.~\cite{Hoekstra02,Percival02}). 
If $\tau=0.04$ and $\Omm=0.3$ then our result is very consistent with the
new lower cluster normalisation found by 
Refs.~\cite{Borgani02,Seljak02,Reiprich02,Viana02}
and just consistent with
the cosmic shear measurements of 
Refs.~\cite{VanW02,Bacon02,Refregier02,Hoekstra02}.
The high clustering amplitude required to fit the small scale
clustering observed by CBI of $\sigma_8\sim 1$ (Ref.~\cite{Bond02})
is in the tail of the distribution and may require
an optical depth rather larger than expected in simple models.

By using subsets of the chains we have checked that the Monte-Carlo sampling noise is negligible at
 the accuracy quoted. The results are Monte-Carlo marginalized over all
of the other parameters and also analytically or numerically
marginalized over the calibration type uncertainties discussed in Section~\ref{cmbonly}.

We now demonstrate the power of the Monte-Carlo method by using the
above data to constrain a larger number of parameters,
using the proposal density described in
Appendix~\ref{MCMC} to exploit the differing computational costs of
changing the various parameters.
We consider separately the case of inflationary models, which
are flat, and more general models less constrained by theoretical prejudice.
In both cases we include $f_\nu$, the fraction of the dark matter that is in the
form of massive neutrinos\footnote{We assume three neutrinos of degenerate
mass, as indicated by the atmospheric and solar neutrino oscillation
observations~\cite{Kamiokande01,Barger02}, and compute the evolution using the fast but accurate method
described in Ref.~\cite{Lewis02}.}, and allow for an effective constant equation
of state parameter\footnote{
Many quintessence models can be described accurately by a constant effective
equation of state parameter~\cite{Huey99}. We compute the perturbations
by using a quintessence potential $V(\phi)$ with $V_{,\phi} = -\numfrac{3}{2}(1-w) H
\dot{\phi}$ and $V_{,\phi\phi} = -\numfrac{3}{2}(1-w)[ \dot{H} -
\numfrac{3}{2}(1+w) H^2 ]$ that gives a constant equation of state.
%AQuintessence does not predict a constant
%equation of state parameter, 
%and in general there are perturbations which havesome effect on the
%large scale CMB anisotropy. 
%one would expect a better fit assuming$w\ne -1$ than fixing $w=-1$
%(equivalent to a cosmological constant). 
%Our parameterization can also be thought of as particular way of
%accounting for more general changes to the late time background
%evolution. Neglecting the perturbations gives our theoretical predictions
% more CMB power on large scales than would be expected in most quintessence models.
} $w\equiv p/\rho$ for the
%A homogeneous 
dark energy, and assume
that $-1 \le w < 0$.

\subsection*{Inflationary models}

The simplest single-field inflationary models predict a flat universe and can be
described quite accurately by the slow-roll approximation. The shape
and amplitude of the initial curvature perturbation depends on the
shape of the inflationary potential, often encoded in `slow-roll
parameters' which are assumed to be small, plus an overall
normalization which depends on the Hubble rate when the modes left the
horizon during inflation. The initial scalar and tensor power
spectra are 
parameterized as usual 
by\footnote{
Here defined so $\la |\chi|^2 \ra = \int \d \ln k \, P_\chi(k)$ and $\la
h_{ij} h^{ij} \ra = \int \d\ln k\, P_h(k)$, where $\chi$ is the initial
curvature perturbation and $h_{ij}$ is the transverse traceless part
of the metric tensor.
These definitions ensure $P=\text{const}$ corresponds to scale invariant.}
\begin{equation}
\label{powerdef}
P_\chi(k) = A_s \left(\frac{k}{k_{s0}}\right)^{n_s-1} \quad\quad
P_h(k) = A_t \left(\frac{k}{k_{t0}}\right)^{n_t}
\end{equation}
where $n_s$ and $n_t$ are the conventional definitions of the spectral indices.
At the lowest order approximation the slow-roll initial power
spectra are determined from the inflationary potential $V$ by the slow-roll parameter parameters
$\epsilon_1$, $\epsilon_2$ by~\cite{Leach02}
\begin{eqnarray}
\epsilon_1 &=& \frac{A_t}{16 A_s} =  \frac{\mpl^2}{16\pi} \left(\frac{V'}{V}\right)^2 \\
n_s &=& 1 - 2\epsilon_1 - \epsilon_2 \equiv  1 - \frac{\mpl^2}{8\pi} \left[ 3 \left(\frac{V'}{V}\right)^2
-\frac{V''}{V} \right] \\
A_s &=& \frac{H^2}{\pi \epsilon_1 \mpl^2} \quad\quad\quad n_t =
-2\epsilon_1 = -\frac{A_t}{8 A_s}
\end{eqnarray}
where quantities are evaluated when $H a= k_\star$ (we use
$k_\star = k_{s0} = k_{t0} = 0.01 \Mpc^{-1}$). For our analysis we use the
parameterization of Equations~\eqref{powerdef} and define $\epsilon_1
\equiv A_t/16 A_s$. We also impose the slow-roll constraint that
spectral index of the tensor modes is given by $n_t =
-2\epsilon_1$. Our results will be consistent with inflationary models
in the region of parameter space in which $\epsilon_1 \ll 1$, $n_s
\approx 1$, but elsewhere can be interpreted more generally (the
results are not very sensitive to the tensor spectral index).
From the definition it is clear that $\epsilon_1 \ge 0$,
and except in contrived models one also expects 
$n_s \le 1$, though we
do not enforce this.  Simple ekpyrotic models are consistent with this
parameterization when there are no tensor modes~\cite{Ekpyrosis}.
If there were evidence for tensor modes ($\epsilon_1>0$) then this 
would be direct evidence against simple ekpyrotic models.

\begin{figure}
\begin{center}
\BW{
\epsfig{figure=inflation_bw.ps,angle=0,width=13cm} 
}{
\epsfig{figure=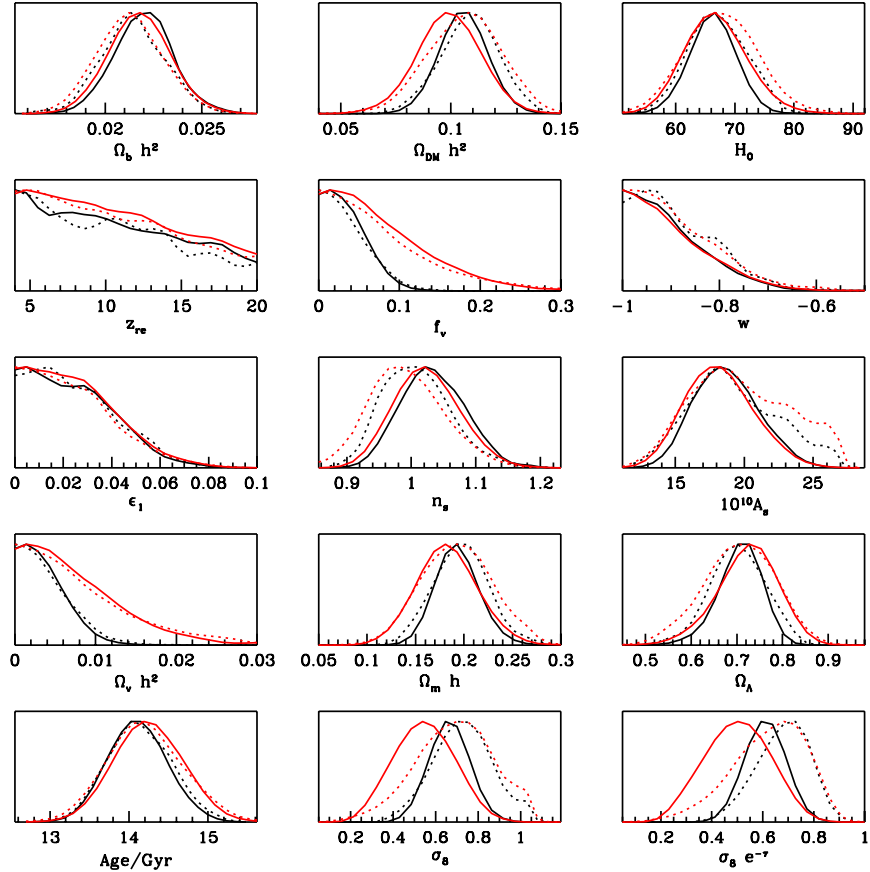,angle=0,width=13cm} 
}
\caption{Posterior constraints for 9-parameter flat models using all data. 
The top nine plots show the constraints on the base MCMC parameters, the remaining plots
show various derived parameter constraints. \BW{Thin}{Red} lines include CMB, HST, SnIA and BBN constraints, \BW{thick}{black} lines also
include the 2dF data. The solid
lines show the fully marginalized posterior, the dotted lines show the
relative mean 
%A posterior
likelihood 
of the samples. The curves are generated
from the MCMC samples using a Gaussian smoothing kernel $1/20$th the width
of each plot.
\label{inflation1D}}
\end{center}
\end{figure}

Figure~\ref{inflation1D} shows the fully marginalized posterior
constraints on the various parameters using the CMB, supernovae,
HST, and nucleosynthesis constraints, with and without the 2dF data,
generated from 7700 weakly correlated samples. 
We generate samples without the 2dF or CBI data, and then importance
sample including CBI to compute results with and without the 2dF data.
The constraints on $\Omm h$ and $f_{\nu}$ are sharpened significantly
on adding in 2dF.
The large shift in the $\sigma_8$ distribution 
comes from the exclusion of the high $f_\nu$ parameter space due to
the new constraint on the shape of the matter power spectrum 
(see discussion of degeneracy below).

\begin{figure}
\begin{center}
\epsfig{figure=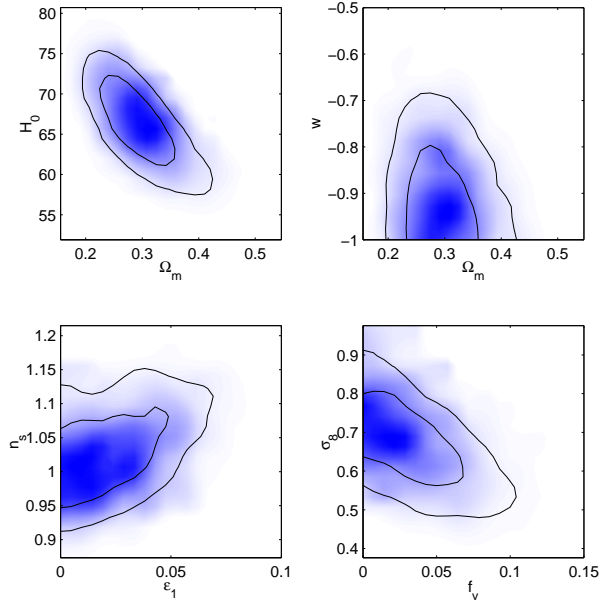,angle=0,width=8cm} 
\caption{All-data posterior constraints for flat inflationary models using. The
contours show the $68\%$ and $95\%$ confidence limits from the
marginalized distribution. The shading shows the mean likelihood of
the samples, and helps to demonstrate where the marginalized
probability is enhanced by a larger parameter space rather than by a
better fit to the data (e.g. low $n_s$ values fit the data better).
\label{inflation2D}}
\end{center}
\end{figure}

It is important to check that the parameters are really
being constrained, in the sense that the results are relatively
insensitive to the priors, so in addition to the marginalized
posterior we also plot the mean likelihood of the samples. These will
differ in general, particularly when the result is sensitive to the parameter space volume
available, which can change as the result of choosing different priors
(see Appendix~\ref{constraints}). 
In
most of the 1-D plots the two methods are in good agreement indicating that
the likelihood is well constrained in n-D space and the priors are
not biasing our results. However the marginalized value of $\sigma_8$
is brought down by the $f_\nu >0$ phase space (since massive neutrinos
damp the small scale power), 
even though the best fits to the data occur where the neutrinos are very
light
(the correlation is shown in the bottom right hand panel of 
Figure~\ref{inflation2D}.)
Similarly the marginalized value of $n_s$ is 
slightly
increased by the phase
space with $\epsilon_1>0$; this increases the CMB power on large
scales, and hence requires a higher spectral index for the scalar modes
(bottom left panel of Figure~\ref{inflation2D}).

We also show in Figure~\ref{inflation2D} that a small degeneracy
between the Hubble constant and the matter density remains (top left)
after the $\sim \Omm h$ constraint from the galaxy power spectrum
shape is combined with the $\sim \Omm h^3$ CMB constraint 
(Table~\ref{cmbflatpars}).
Geometrical information from the CMB peak position and the SnIA
work together to constrain $w$, but this remains slightly correlated
with $\Omm$ (top right).

%%%%%%%%%%%%%%%%%%%%%%%%%%%%%%%%%%
%%%%%%%%%%%%%%%%%%%%%%%%%%%%%%%%%%

%%%%%%%%%%%%%%%%%%%%%%%%%%%%%%%%%%
%%%%%%%%%%%%%%%%%%%%%%%%%%%%%%%%%%

The results from the 6 and 9 parameter analyses can be compared 
using the 68 per cent limits given in Table~\ref{inflationtable}
and the 
plots
in Figure~\ref{full1D}. 
Many of the results are quite robust to the addition of the extra
three
degrees of freedom.
The biggest change is in $\sigma_8 e^{-\tau}$ which
is brought down by contributions from non-zero $f_{\nu}$.
%A these numbers should now all be updated. 
\begin{table}
\vbox{\center{\centerline{
\begin{tabular}{|c||c||c|c|c|c|}
\hline
   & 6 parameters & \multicolumn{4}{|c|}{9 parameters}\\
\cline{2-6}
               & +2dF & no 2dF & +2dF    & +2dF & +2dF  \\
\cline{2-6}
               & $68\%$-$1$D  & $68\%$-$1$D &$68\%$-$1$D  & $68\%$-full    & $95\%$-full  \\
\hline 
$\bm{f_v}$          & - & $<0.10$ & $<0.04$          &   $<0.10$    &  $<0.13$\\
$\bm{w}$            & - & $<-0.87$ &$< -0.88$        & $< -0.68$   & $< -0.58$ \\
$\bm{\epsilon_1}$   & - & $<0.032$& $< 0.032$   & $< 0.069$      & $< 0.085$      \\
$m_\nu/\!\eV$         & - & $<0.29$&  $ <0.14$     & $< 0.36$  & $<0.54$ \\ 
$r_{10}$     & - & $<0.30$ & $< 0.31$        &$<0.92$    &    $<1.4$  \\
\hline
$\bm{\Omb h^2}$     & $0.021\pm 0.001$ &$0.022\pm 0.001$&$0.022\pm 0.001$       & $0.018-0.025$ & $0.017-0.026$    \\ 
$\bm{\Omdm h^2}$    & $0.113\pm0.008$ &$0.099\pm 0.014$& $0.106\pm 0.010$     & $0.082-0.130$ & $0.072-0.142$\\
$\bm{h}$            & $0.67\pm 0.03$ &$0.67\pm 0.05$& $0.66\pm 0.03$          & $0.59-0.75$   & $0.55- 0.78$\\
$\bm{n_s}$          & $0.98 \pm 0.04$ &$1.02\pm 0.05$&  $1.03\pm 0.05$     & $0.91-1.13$  & $0.87-1.19$ \\ 
$\Oml$       & $0.70\pm 0.04$ &$0.72\pm 0.06$&  $0.71\pm 0.04$       & $0.58-0.80$   & $0.54-0.82$   \\
$\Omm$       & $0.30\pm 0.04$ &$0.28\pm 0.05$&  $0.29\pm 0.04$       & $0.20-0.42$   & $0.18-0.46$   \\
$t_0/\Gyr$   & $14.1\pm 0.4$  &$14.3\pm 0.4$&  $14.1\pm 0.4$       & $13.3-15.0$   & $ 13.0-15.2$ \\
$\Omm h$    & $0.20\pm 0.02$  &$0.18\pm 0.03$& $0.19\pm 0.02$   & $0.15-0.25$  &  $0.13 - 0.26$\\
$\sigma_8$  & $0.79 \pm 0.06$ &$0.54\pm 0.13$& $0.67\pm 0.08$ & $0.49-0.93$ & $0.45- 0.95$ \\
$\sigma_8 e^{-\tau}$ & $0.72\pm 0.04$ &$0.50\pm 0.12$&  $0.61\pm 0.07$        &  $0.47-0.81$  & $0.41-0.84$  \\ 
$\sigma_8 \Omm^{0.55}$ & $0.40 \pm 0.05$ &$0.27\pm 0.08$& $0.34 \pm 0.05$  & $ 0.22-0.51$ & $0.19 - 0.53$ \\
\hline
\end{tabular}
}}}
\caption{
Parameter constraints for 6 and 9 parameter flat models with all data
%A
with or without 2dF. The top section shows the
constraints on the additional parameters that were fixed
in the basic 6 parameter model, the bottom half shows the effect these
additional parameters have on the results for the basic
parameters. 1D limits are from the confidence interval of the
fully marginalized 1D distribution, the full limits give the extremal
values of the parameters in the full n-dimensional confidence
region (see Appendix~\ref{constraints} for discussion). 
Bold parameters are base Monte-Carlo parameters, non-bold parameters
are derived from the base parameters.
}
\label{inflationtable}
\end{table}

As discussed in Appendix~\ref{constraints}, 
parameter confidence limits from the full 
n-D
distribution can also 
easily be
calculated from a list of samples.
We show the marginalized and n-D
parameter
constraints with the inflationary assumptions in
Table~\ref{inflationtable}.\footnote{
Monte-Carlo samples from the posterior do
not provide accurate estimates of the parameter best-fit values (in
high dimensions the best-fit region typically has a much
higher likelihood than
the mean, but it occupies a minuscule fraction of parameter space)
therefore we do not quote best fit points.
The high-significance limits 
are also hard to calculate
due to the scarcity of samples in these
regions. To compute accurate estimates in the tails of the
distribution
and to ensure the tails are well explored,
we sample from a broader distribution and then importance sample to the
correct distribution,  by originally sampling from
$P^{1/T}$ where $T>1$ (we use $T=1.3$). 
}
As expected, the n-D limits are much wider than those from the
marginalized distributions, most being more than twice as wide. 

The combined datasets provide good constraints on the neutrino
mass, despite the large parameter space. The massive neutrino fraction
$f_\nu$ translates into the neutrino mass via
\begin{equation}
\Omnu h^2 = f_\nu \Omdm h^2 = \frac{\sum m_\nu}{93.8\eV} \quad\quad \implies
\quad\quad m_\nu \approx 31 \, \Omnu h^2 \eV,
\end{equation}
where the last equality follows from our assumption that there are
three neutrinos of approximately degenerate mass, as indicated by the
small mass-squared differences detected by the neutrino
oscillation experiments~\cite{Kamiokande01,Barger02}. 
At $95\%$ confidence we
find the marginalized result $m_\nu \alt 0.27 \eV$ and the more
conservative n-D result $m_\nu \alt 0.5\eV$. 
The tightness of the constraint is predominantly due to the 2dF data, 
as shown in Figure~\ref{inflation1D}, via the damping effect of
massive neutrinos on the shape of the small scale matter power spectrum.\footnote{
Our strategy of generating chains without the 2dF data and then importance
sampling ensures that we have many samples in the tail of the
distribution, and hence that our upper limit is robust 
(since we have have much lower Monte-Carlo noise in the tails than if we had
generated chains including the 2dF data), and also makes the effect of
the 2dF data easy to assess. }
The result is consistent with the
weaker limits found in
Refs.~\cite{Elgaroy02, Hannestad02} under more restricted assumptions.
The marginalized result is only slightly affected by the large parameter
space: computing chains with $w=-1$ and $A_t=0$ we obtain the
marginalized $95\%$ confidence result $m_\nu \alt 0.30\eV$ (the n-D
limit is much less sensitive to the parameter space, and the result
does not change significantly).
Thus the simplest model where all the neutrino masses are very small is still a good
bet.

%A
The result for the quintessence parameter $w$ is consistent
with $w=-1$, corresponding to a cosmological constant. The
marginalized limit is $w<-0.75$ at $95\%$ confidence, consistent with
%A finally keep Alessandro happy!
Ref.~\cite{Bean01}.
If we neglect the quintessence perturbations it is a simple matter to
relax the assumption that $w\ge -1$; for flat models with no tensors or massive neutrinos
we find the marginalized result $-1.6< w <-0.73$ at $95\%$
confidence, and the n-D result $-2.6 < w < -0.6$ with the best fit close to
$w=-1$, broadly consistent with Ref.~\cite{Hannestad02;2}. 
%A
Note that including quintessence perturbations leads to a tighter
constraint on $w$ due to the increased large scale power. Although
perturbations are required for consistency with General Relativity,
it is possible that a quintessence model may be approximated better by a constant $w$
model neglecting perturbations than one including the perturbations.

The constraint on the tensor mode amplitude (encoded by $\epsilon_1$) is weak, as expected due to the
large cosmic variance on large scales. In
table~\ref{inflationtable} we also show the result for $r_{10} \equiv C_{10}^T/C_{10}^S$, the
ratio of the large scale CMB power in tensor and scalar modes. For comparison, with perfect
knowledge of all the other parameters and a noise-free sky map, the CMB temperature power
spectrum cosmic variance detection limit is $r_{10}\agt 0.1$.

The method we have used could be generalized for a more accurate
parameterization of the initial power spectrum, for example going to
second order in the slow-roll parameters~\cite{Leach02}, which in
general introduces a running in the spectral index. The current
data is however clearly consistent with the simplest scale invariant power
spectrum with no tensor modes. 
As a check we have generated chains for
flat models with a cosmological constant, no massive neutrinos or
tensors, but allowing for a running spectral index, and found the
$68\%$-confidence marginalized result 
$-0.06 < \nrun < 0.02$ at $k=k_0=0.05\Mpc^{-1}$
where $\nrun \equiv d^2 (\ln P_\chi)/d (\ln k)^2$.
This corresponds to the running spectral index
$n_{s,\text{eff}}(k) \equiv d \ln P_\chi/d\ln k = n_s(k_0) + \nrun \ln (k/k_0)$.

\subsection*{Non-flat 11-parameter models}

We now relax the constraint on the curvature, and allow the tensor
spectral index to be a free parameter (we assume $n_t\le 0.1$). We
parameterize the power spectra of the initial curvature and tensor
metric perturbations as in the inflationary case\footnote{For non-flat models our definitions
follow~\cite{Gratton01}. In open models we assume the tensor power
spectrum has an additional factor of $\tanh(\pi\sqrt{-k^2/K-3}/2)$ on
the right hand side.}, except that we now report results for
$A_t/A_s$ rather than a slow-roll parameter, and choose the scalar and
tensor pivot scales $k_{s0} = 0.05 \Mpc^{-1}$, $k_{t0} = 0.002 \Mpc^{-1}$
(the standard \CMBFAST\ parameterization).

\begin{figure}
\begin{center}
\BW{
\epsfig{figure=full_comp_bw.ps,angle=0,width=13cm} 
}{
\epsfig{figure=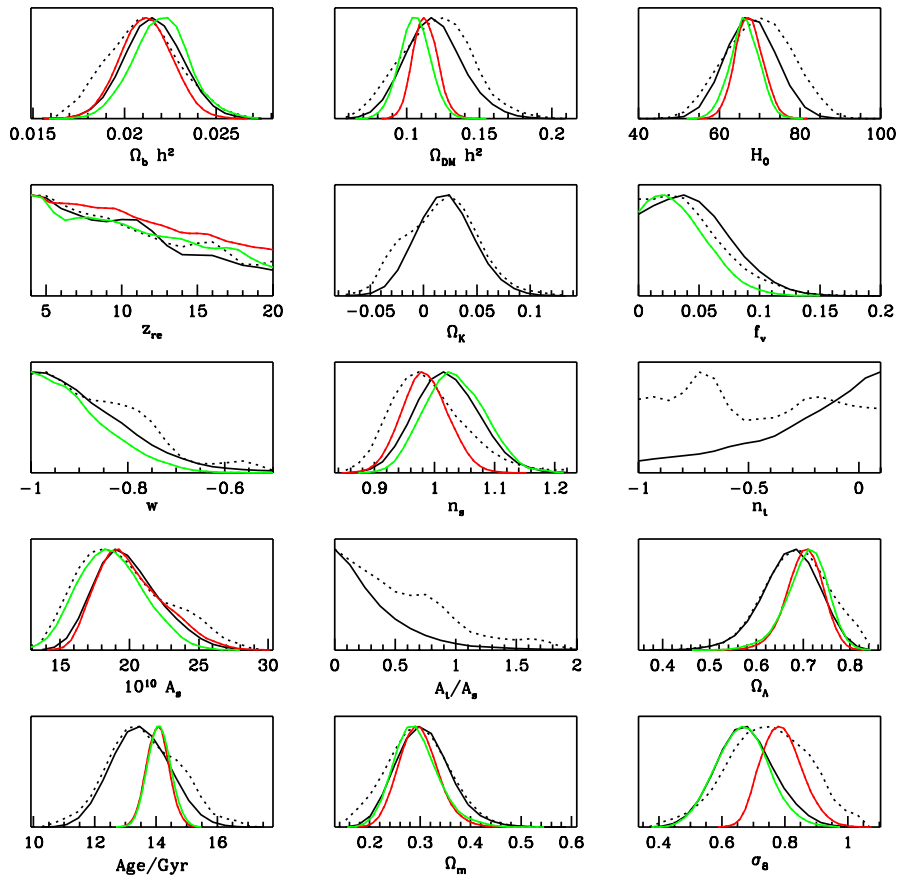,angle=0,width=13cm} 
}
\caption{
Posterior constraints for $11$-parameter non-flat
models (\BW{thick lines}{black lines}) using all data, compared with
$6$ \BW{(thin, solid lines)}{(red)} and $9$ \BW{(thin, dashed
lines)}{(green)} parameter models.
Dotted lines show the mean likelihood of the samples for the
$11$-parameter model. 
Some sampling noise is apparent due to the relatively small number of
samples used. 
\label{full1D}}
\end{center}
\end{figure}

In Figure~\ref{full1D} we show the parameter constraints that we get
using about 10000 weakly correlated samples
importance sampled to include the 2dF data and CBI. 
For comparison we plot the equivalent constraints with the $9$ and
$11$ parameter models. The additional freedom in the curvature
broadens some of the constraints significantly, though the $\Omm$ and
$\Omb h^2$ constraints are quite robust.

The tensor spectral index is
essentially unconstrained, the difference between the mean likelihood
and marginalized 1D plots being due to the assumed flat prior on the
tensor amplitude --- at very small amplitudes $n_t$ could be
anything and still be undetectable. 
%A updates for w perturbations
%There are now two additional parameters affecting the late time evolution, $w$
%and $\OmK$, and as such the constraints on both are considerably weakened; 
The $95\%$ marginalized limit on the curvature is $-0.02 <\OmK <0.07$.
%A$-0.02 <\OmK < 0.09$, $w<-0.55$.
%Open models with $w\sim -0.8$ are slightly favored, though as shown in Figure~\ref{full2D} 
%a flat universe with $w=-1$ is within the $68\%$ confidence contour .
Slightly open models fit the data marginally better on average, though
a flat universe is well within the $68\%$ confidence contour.
The limit on the equation of state parameter is slightly weakened to
$w<-0.69$, and neutrino mass is now $m_\nu < 0.4\eV$ at $95\%$
confidence.

%%%%%%%%%%%%%%%%%%%%%%%%%%%%%%%%%%
%%%%%%%%%%%%%%%%%%%%%%%%%%%%%%%%%%

%%%%%%%%%%%%%%%%%%%%%%%%%%%%%%%%%%
\begin{figure}
\begin{center}
\epsfig{figure=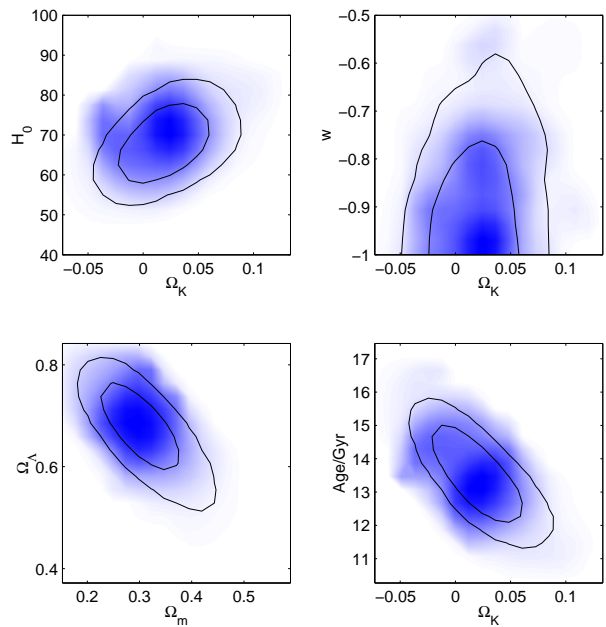,angle=0,width=8cm} 
\caption{Posterior constraints for 11-parameter non-flat models using
all data. 
%AFlat models with $w=-1$, $\OmK = 0$ are within the inner
%$68\%$-confidence contours, though the mean likelihood (shading) is very
%slightly higher for open models with $w\sim -0.8$, $\OmK \sim 0.025$.
\label{full2D}}
\end{center}
\end{figure}
%%%%%%%%%%%%%%%%%%%%%%%%%%%%%%%%%%

\subsection*{Which model fits best?}

We have explored the posterior distribution in various parameter
spaces, deriving 
parameter constraints in the different models. 
Since we obtain only upper limits on $f_{\nu}$, $w$ and $A_t/A_s$
there is no 
evidence for massive neutrinos, $w\ne -1$ or tensor modes
using current data. 

One can make a more quantitative comparison of the different models by
comparing how well each fits the data. As discussed in
Appendix~\ref{constraints} a natural measure is the mean likelihood of
the data obtained for the different models. Equivalently, if one chose
a random sample from the possible parameter values, on average how
well would it fit the data? 
We find that the six, nine and eleven parameter models have mean
likelihoods ratios 
$1:0.4:0.3$ using all the data.
So by moving away from the basic model we have not increased the goodness of
fit on average (rather the reverse), which is not surprising given
how well the basic model fits the data. Most of the 
distributions of the additional parameters peak at their fixed values.

We also considered the probability of each model, found from 
marginalizing out \emph{all} parameters (the `Evidence' as 
e.g. explained in Ref.~\cite{Sivia96}). 
Since the 9 and 11 parameter models are totally consistent 
with the 6 parameter model then it is already clear that using
this method will favor the 6 parameter model 
for any choice of prior.
The numerical evidence ratio depends very strongly on the prior, and
without a well motivated alternative to the null hypothesis (that there are only 6
varying parameters), its value is not useful. 
The mean likelihood of the samples (above) uses the posterior
as the prior, which is at least not subjective,
and has a convenient interpretation in terms of goodness of fit.

 Whilst certainly not ruled out, at the moment there is no
evidence for observable effects from the more complicated models we
have considered. Nonetheless, when considering parameter values, it is
important to assess how dependent these are on the assumptions, and
this can be seen by comparing the results we have
presented. We conclude that at the moment simple inflationary models with
small tilt and tensor amplitude (e.g. small field models with a nearly flat
potential; or observationally equivalently, ekpyrotic models) account
for the data well. On average the fit to the data is not improved by adding
 a cosmologically
interesting neutrino mass or by allowing the dark energy to be 
something other than a cosmological constant.

\section{Conclusions}

In this paper we have demonstrated the following benefits of sampling 
methods for cosmological parameter estimation:
\begin{itemize}
\item
The practicality of exploring the full shape of high-dimensional posterior
parameter distributions using MCMC.
\item
The use of principle component analysis to identify well
constrained non-linear combinations of parameters and identify
degeneracies.
\item
Simple calculation of constraints on any parameters that can be
derived from the base set (e.g. age of the universe, $\sigma_8$, $r_{10}$, etc.).
\item
Use of the mean likelihood of the samples as an alternative to
marginalization to check robustness of results
and 
relative goodness of fit.
\item 
The calculation of extremal values within the n-D hyper-surface 
to better represent the range of 
the full probability distribution.
\item
The use of importance sampling to quickly compare results with
different subsets of the data, inclusion of new data, and correction for
small theoretical effects.
\end{itemize}
Our Monte-Carlo code and chains are publicly available at 
\url{http://cosmologist.info/cosmomc}.

With the current cosmological data we found that the
Monte-Carlo approach works well, though 
simply picking the best fit sample does not identify best-fit model to 
high accuracy
(and therefore we do not quote these numbers), 
and there are potential difficulties
investigating posterior distributions with multiple local minima 
(although this is not a problem given the parameters and data 
used here).

We investigated a 6-D cosmological 
parameter space and found, for the first time, a concise description
of the CMB constraints on the matter power spectrum normalization,
in addition to tight constraints on $\Omb$ and $n_s$ in agreement
with previous analyses. The new information from the CBI and VSA
interferometers is in good agreement with the older data points and
we find that our results are negligibly changed on removing 
this information.

On adding in constraints from a wide range of cosmological
data we evaluated constraints on the above 6 parameter model as well
as extending to more complicated 9 and 11 parameter models.
Many of
the constraints on the base set of parameters were 
fairly robust to the addition of this extra freedom, for example
the matter density changed 
from $\Omm=0.30 \pm 0.04$ for the basic 6 parameter model to 
$\Omm=0.28 \pm 0.07$ for the 11 parameter model (68 per cent confidence).
On the other hand the value for the matter power spectrum
normalization on $8 h^{-1} {\rm Mpc}$ scales
is quite dependent on 
the neutrino mass,
and allowing for a significant neutrino mass decreases the mean value of
$\sigma_8$ 
(the constraint on the amplitude could be improved by
better constraints on the small scale CMB amplitude and 
the reionization redshift.) 
Parameters affecting or sensitive to the late time evolution tend to be rather
degenerate, and constraints on these are considerably weakened on
adding additional freedom in the model.

We find that the 9 parameter model is quite well constrained by the
amount of data used and obtain upper limits on a number of 
interesting cosmological parameters, given our assumptions of
a flat universe with slow-roll inflation constraints. 
In particular we find the marginalized constraint $m_\nu \alt 0.3\eV$
on the neutrino mass and $w \alt -0.75$ for the
equation of state parameter (95 per cent confidence). There is no
evidence for tensor modes, though the constraint is currently quite
weak, with the constraint on the ratio of the large scale CMB power
being $r_{10} \alt 0.7$.
This constraint could be sharpened considerably by restricting the
allowed range of scalar spectral indices and neutrino masses.

%A
In the 11 parameter space the limits are weakened slightly and the
standard cosmology of $w=-1$ and $\OmK=0$ is near the peak of the
posterior probability.
%consistent 
%In the 11 parameter space there is a degeneracy between 
%between the quintessence
%parameter $w$ and the curvature $\OmK$ (Figure~\ref{full2D}).
%However the standard cosmology of $w=-1$ and $\OmK=0$ is still
%consistent and the degeneracy is slightly broken giving $w<-0.54$.
The tensor spectral index is essentially unconstrained 
as expected given that the only information comes from the large
scale (COBE) CMB data.

While a detailed investigation of the effect of using all the
different combinations of cosmological constraints is beyond the
scope of this paper we do show the effect of removing the second
most powerful constraint (the galaxy power spectrum) on the 9 
parameter model in Figure~\ref{inflation1D}. The limits on 
most of the parameters are affected remarkably little. 
The neutrino mass is the most affected, with the upper limit 
doubling on removing 2dF.
The neutrino mass is 
correlated
with the 
matter power spectrum shape parameter (roughly $\Omm h$)
and amplitude, and these constraints are correspondingly
weakened on removing 2dF.

As new better data become available our
general method should also be
applicable into the future. Due to the enormously decreased number of
likelihood evaluations in the MCMC method compared to other approaches,
theoretical predictions 
can be computed essentially exactly, and one can account for the available data in detail.

\begin{acknowledgments}
We thank the members of the Leverhulme collaboration for many useful
discussions, in particular Ofer Lahav, Carolina \"Odman, Oystein
Elgaroy, Jerry Ostriker and Jochen Weller.
We are grateful to David MacKay and Steve Gull for encouraging the use of
MCMC.
We thank Anze Slozar, Keith Grainge, Alexandre R\'efr\'egier 
and Kev Abazajian
for helpful suggestions.
We thank George Efstathiou for making his CMBfit code
available to us.
AL thanks the Leverhulme Trust for support.
SLB acknowledges support from Selwyn College and PPARC and thanks 
the Aspen Center for Physics where part of this work was done.
 We thank PPARC and HEFCE for support of the COSMOS facility. 
\end{acknowledgments}

\appendix
\section{The Metropolis-Hastings algorithm}
\label{MCMC}

The algorithm that we use for generating samples from the posterior
distribution using a Markov Chain is 
the Metropolis-Hastings algorithm.
For an introduction and overview of MCMC methods see Ref.~\cite{MCMC97,MacKayBook,Neil93}.
A Markov Chain moves from a position in parameter space $\vtheta_1$ to
the
next position $\vtheta_2$ with transition
probability $T(\vtheta_1, \vtheta_2)$, where $\vtheta$ labels a vector of
parameter values. The Metropolis-Hastings transition kernel $T(\vtheta_1, \vtheta_2)$ is chosen
so that the Markov Chain has a stationary asymptotic distribution
equal to $P(\vtheta)$, where $P(\vtheta)$ is the distribution we wish to
sample from. This is done by using an arbitrary {\it proposal density} distribution
$q(\vtheta_n,\vtheta_{n+1})$ to propose a new point $\vtheta_{n+1}$ given
the chain is currently at $\vtheta_n$. The proposed new point is then accepted  with probability
\begin{equation}
\alpha(\vtheta_n,\vtheta_{n+1}) = 
\min\left\{ 1, \frac{P(\vtheta_{n+1})q(\vtheta_{n+1},\vtheta_{n})}
{P(\vtheta_{n}) q(\vtheta_n,\vtheta_{n+1})} \right\}
\end{equation}
so that
$T(\vtheta_n,\vtheta_{n+1})=\alpha(\vtheta_n,\vtheta_{n+1})q(\vtheta_n,\vtheta_{n+1})$.
 This construction ensures that detailed balance holds, 
\begin{equation}
P(\vtheta_{n+1}) T(\vtheta_{n+1},\vtheta_n) =P(\vtheta_{n}) T(\vtheta_{n},\vtheta_{n+1}), 
\end{equation}
and hence that $P(\vtheta)$ is the equilibrium distribution of the
chain. 

If the chain is started in a random position in parameter space it
will take a little time, {\it burn in}, to equilibrate before it starts sampling from the posterior
distribution. After that time each chain position is a {\it correlated}
sample from the posterior. The correlation is particularly obvious if
the proposal is not accepted as then there are two or more samples at
exactly the same point. 
However by 
using only occasional chain positions
(thinning the chain) one can give 
the chain time to move to an uncorrelated position in
parameter space, and 
independent samples are then obtained. 
Small
residual correlations between samples are unimportant for almost all
calculations, though they do make the Monte-Carlo error on the results
harder to assess.

%A comment(1)
For the cases we consider the chains equilibrate rapidly, at worst
after thousand or so points. The results can be checked easily
by using a longer burn in and comparing results. We thin the chain positions by
a factor of 25-50 depending on the number of parameters, leaving
weakly correlated samples that we use for importance sampling
(see Appendix~\ref{importance}).

If the proposal density is symmetrical it cancels out when working out
the acceptance probability, which then becomes just the ratio of the
posteriors. This is the case when 
the proposal density is independent of the current position of the chain, which is the case we consider.

\subsection*{The proposal density}

The choice of 
proposal density can have a large effect on how the
algorithm performs in practice. In general 
it is best to have a
proposal density that is 
of similar shape to
the posterior, since this ensures
that large changes are proposed to parameters along the degeneracy directions.
Fortunately with cosmological data we have a 
reasonable idea of what the posterior might
look like, and so choosing a sensible proposal density
is not difficult.

If posteriors from models with common parameters are much easier to compute it can be
 beneficial to use a proposal density that changes only a subset
of the parameters on each iteration, ensuring that consecutive
posterior evaluations only differ in a subset of the
parameters. Proposing a change to a random subset of the parameters
also increases the acceptance rate, especially in high dimensions,
giving faster piecewise movement around parameter space. 
In the case of CMB parameter estimation,
models that differ only by a different normalization of the
theoretical CMB power spectrum are very quick to compute
once the $C_l$ values for a single model have been calculated.
Similarly changing parameters that
govern calibration uncertainties in the data can also be very
quick. However changing parameters that govern the perturbation
evolution, for example $\Omb$, $\Omc$, etc, will be much
slower as in general it requires a detailed recalculation of the linear physics.

If we are comparing CMB data with theoretical models, the most general
way to compute the theoretical $C_l$ power spectrum is using a
fast Boltzmann code such as \CAMB~\cite{Lewis99} (a
parallelized version of \CMBFAST~\cite{Seljak96}; we discuss less accurate and general schemes below). Since the perturbation evolution is assumed to be linear, any
parameters governing the initial power spectra of the scalar and
tensor perturbations will be fast to compute once the transfer
function for each wavenumber has been computed. Parameters governing
the initial power spectrum are therefore `fast' parameters.

We therefore use a proposal density that makes changes only within the
subsets of the fast and slow parameters, at least when we do not have
an approximate covariance matrix available for the posterior.\footnote{When changing the
slow parameters it is possible to also change the fast
parameters at the same time. This can be a good idea when there are highly correlated
slow and fast parameters, for example the reionization redshift and
the tensor amplitude.}
We made the fairly arbitrary choice to change a subset of one to three parameters at a
time, cycling through the parameters to be changed in random order,
which gives a high acceptance rate ($\sim 50\%$) for the cases
%A we have answered part of (1) here already
we considered.
After one initial run
one can transform to a set of parameters which diagonalize the 
covariance matrix before doing subsequent runs, allowing efficient
exploitation of degeneracy information as long as the posterior is
reasonably Gaussian.

The above scheme is sufficient for parameter estimation from current
data, however as more data becomes available 
the posterior
may become
highly non-Gaussian or disjoint,
in which case 
it may become necessary
to use more
sophisticated schemes using simulated annealing, hybrid Monte Carlo,
or schemes using cross-chain
information~\cite{MCMC97,MacKayBook,Neil93}.
However when the posterior is not disjoint one can often transform to a set
of base parameters which are relatively independent, in which case a
simple Monte-Carlo scheme should continue to work well (see
Appendix~\ref{constraints} for further discussion).

\section{Importance sampling}
\label{importance}
Given a set of samples from a distribution $P$, one can estimate
quantities with respect to a different similar distribution $P'$,
by weighting the samples in proportion to the probability ratios. This
effectively gives a collection of non-integer weighted samples for
computing Monte-Carlo estimates. For example the expected value of a
function $f(\vtheta)$ under $P'$ is given by
\begin{eqnarray}
\left\langle f(\vtheta) \right\rangle_{P'} &=& \int \d\vtheta P'(\vtheta) f(\vtheta) =
\int \d\vtheta \frac{P'(\vtheta)}{P(\vtheta)} P(\vtheta) f(\vtheta) \nonumber\\
&=&\left\langle \frac{P'(\vtheta)}{P(\vtheta)} f(\vtheta) \right\rangle_{P}.
\end{eqnarray}
Given a set $\{\vtheta_n\}$ of $N$ samples from $P$ 
a
Monte-Carlo estimate is therefore
\begin{equation}
\left\langle f(\vtheta) \right\rangle_{P'} \approx \frac{1}{N}\sum_{n=1}^N \frac{P'(\vtheta_n)}{P(\vtheta_n)} f(\vtheta_n).
\end{equation}
For this to work it is essential that $P/P'$ is never
very small, and for a good estimate without massively oversampling
from $P$ one needs $P'/P \sim \text{constant}$ everywhere where $P'$ is significantly
non-zero. If $P'$ is non-zero over only a very small region compared
to $P$ it will be necessary to proportionately oversample from
$P$.

If the distributions are not normalized, so that $\int \d\vtheta
P(\vtheta) = Z$, 
the ratio of the normalizing constants 
can be estimated using
\begin{equation}
\frac{Z'}{Z} = \left\la \frac{P(\vtheta)'}{P(\vtheta)}\right\ra_P
\approx \frac{1}{N} \sum_{n=1}^N \frac{P'(\vtheta_n)}{P(\vtheta_n)},
\end{equation}
and hence
\begin{equation}
\left\langle f(\vtheta) \right\rangle_{P'} \approx \frac{\sum_{n=1}^N
 P'(\vtheta_n)/P(\vtheta_n) f(\vtheta_n)}{\sum_{n=1}^N P'(\vtheta_n)/P(\vtheta_n)}.
\end{equation}

In Bayesian analysis it can be useful to compute the ratio of the
evidences $P(D) = \int \d\vtheta P(D, \vtheta)$, given as above by
\begin{equation}
\frac{P'(D)}{P(D)} = \left\langle \frac{P'(\vtheta,D)}{P(\vtheta,D)}\right\rangle_{P(\vtheta|D)}
\approx \frac{1}{N} \sum_{n=1}^N
\frac{P'(D|\vtheta_n)P'(\vtheta_n)}{P(D|\vtheta_n) P(\vtheta_n)},
\label{evid_ratio}
\end{equation}
where the samples $\{\vtheta_n\}$ are drawn from $P(\vtheta|D)$.
Assuming the distributions are sufficiently similar, the evidence
under $P'$ can therefore easily be computed from the probability
ratios at a sample of points under $P$, and a known evidence
under $P$.  In many cases only the ratio is of interest --- the ratio is
larger than one if on average the probability of the samples under
$P'$ is higher than under $P$. In the case where the distributions are
very different one may need to introduce a series of intermediate
distributions that are all not too dissimilar to each other, and
perform Monte Carlo simulations for each. The
evidence ratio one requires is then just the product of that for all
the intermediate distributions. Many more general schemes are
described in~\cite{Neil93,Neil01}, though in this paper we only consider
importance sampling to similar or subset distributions.

The simplest application of importance sampling is to adjust results
for different priors. For example if one computes a chain with flat
priors on the parameters, one may wish to importance sample to several different
distributions with different priors on various parameters. This will
work well as long as the prior does not skew the distribution too
much or give non-zero weight to only a very small fraction of the models.

\subsection*{Faster Monte-Carlo}

MCMC runs produce correlated samples from the probability
distribution. To obtain independent samples one thins out the chain by
a sufficiently large factor that the chain has had time to move to a
randomly different point between the thinned samples. Depending on how
one implements the MCMC, the shape of the posterior and the number of
dimensions the thinning factor can be quite large,
typically of the order ten to a thousand.

By performing Monte-Carlo simulations with a good approximation to the
true probability distribution one can use importance sampling to
correct the results with an accurate calculation of the
probabilities. This can be useful if computing the probabilities
accurately is much slower than computing an approximation, since one
only ever importance samples {\it independent} samples.
The burn-in and random walking stages of the Monte-Carlo involve a
much larger number of probability evaluations, so 
using a fast approximation when generating the chain 
saves a lot of time.

Calculating the posterior from CMB data requires a calculation of the
theoretical CMB power spectra, $C_l$. Using accurate codes like \CAMB\
and \CMBFAST\ is typically {\it much} slower than computing the
likelihoods from the data once the $C_l$ are known (assuming one uses a
radical data-compression scheme, e.g. see Ref.~\cite{Bond98}). In the
not so distant future we will require to high accuracy $C_l$ up to
$l\sim 2500$, including second order effects such as lensing, and also
the matter power spectrum at various redshifts.  
Without access to a fast supercomputer this may be prohibitive.

With a small number of parameters it is possible to use a grid of models and
interpolate to generate accurate $C_l$ quickly, however as the number
of parameters grows the computational cost of computing the grid grows
exponentially. Also, as second order effects such as gravitational lensing
become important, fast grid generation schemes such as the
$k$-splitting scheme of Ref.~\cite{Tegmark01} become much more difficult to
implement accurately. However these may still be useful as a fast
approximation, 
as long as the independent samples are corrected with a more accurate calculation.
Ref.~\cite{Kosowsky02} describe a scheme for generating $C_l$s very
quickly from a small number of base models, a set of optimized
parameters, and an accurate calculation of how the $C_l$ vary with
these parameters. This gives a very fast approximator over a
restricted range of parameters that may prove useful combined with
importance sampling correction.

It is also possible to use fast semi-analytic schemes. Typically these
are based on a smallish grid of base models, from which the $C_l$s in
general models are computed quickly on the fly by accounting for
changes in the angular diameter distance to last scattering, differing
perturbation growth rates, etc. These approximate schemes can be made quite accurate at 
at small scales, with significant errors mainly at low $l$, precisely where the cosmic variance is large. So
whilst an approximate scheme may produces small systematic errors in
the likelihood, if the error is of the same order as the cosmic
variance or less, the probabilities given the data are bound to be sufficiently
similar for importance sampling to be valid. 

A particular approximate $C_l$ generator we have tried is
CMBfit~\cite{Efstathiou02}, which uses of combination
of base $C_l$ grids and analytic fits. This achieves quite good few percent level
accuracy at high $l$, though larger systematic errors at low
$l$. However the code is fast, and we found that importance
sampling the results with an exact calculation of the $C_l$ gives good
results, and removes systematic biases introduced by the low $l$
approximations. Such an approach can be generalized for more general
late time evolution, for example models with quintessence where the
effect on small scales is due almost entirely to changes in the background
equation of state.

An alternative scheme based on grids of the transfer functions for
each wavenumber can produce more accurate results, like the recently
released DASH~\cite{Kaplinghat02}.
However this is not much faster than generating the $C_l$s exactly
using \CAMB\ on a fast multi-processor machine, and relies on a large
pre-computed grid (which introduces its own limitations). The only
real advantage over CMBfit is that more general initial power
spectrum parameterization could be accounted for easily --- something
that is impossible with schemes based on grids of $C_l$s.

Even without a fast semi-analytic scheme, there are a variety of small
corrections that can be applied post hoc. For example lensing 
affects 
the CMB $C_l$ at the few percent level, so one may wish to compute
chains without including the lensing, then importance sample to correct
the results using an accurate calculation including the
lensing\footnote{However if one is also computing the matter power
spectrum numerically the additional cost of including the lensing
effect is small. 
We have checked that the lensing correction to the results we present
is much smaller than the errors (the lensed power spectra can be
computed with \CAMB\ using
the harmonic approach of Ref.~\cite{Hu00}).
}.
For
small scales at high precision one may also wish to run \CAMB\ at a
high-accuracy setting to check that numerical errors in the default
output are not affecting the results.  Also chains could be generated to lower $l$ and
the effect of the high-$l$ constraints accounted for by
importance sampling. For example we generated the chains using
$l_{\text{max}}=1300$, and then for the nearly independent samples re-computed the power spectra up to
$l_{\text{max}}=2000$ for importance sampling with the CBI data.

Similar methods could be applied for the matter power spectrum using
approximate fittings, see e.g. Refs.~\cite{Eis99,Tegmark01}. However when a fast multi-processor machine
is available, and one is interested in a very large number of
parameters, it is much simpler to 
%Agenerate the entire chain using
Monte-Carlo using
\CAMB\ to generate the CMB power spectra and matter power spectrum,
which is what we did for the results we present. The great advantage
of this approach is that it generalizes trivially if one wishes to
include changes in the physics, for example different quintessence
models, or changes in the initial power spectrum.

\subsection*{Constraints with new data}

Assuming that one has some new data which is broadly consistent with
the current data, in the sense that the posterior only shrinks, one
can use importance sampling to quickly compute a new posterior
including the new data. We have made our MCMC chains publicly
available, so these can be used to rapidly compute new posteriors from
new data without incurring any of the considerable computational cost
of generating the original chain. For example if you 
have a new constraint on $\sigma_8$, you just need
to loop over the samples adjusting the
weights of the samples proportional to the likelihood under the new constraint.
Using importance sampling has the added benefit of
making it very easy to assess how the new data is changing the posterior.

\section{Parameter Constraints}
\label{constraints}

The great advantage of the Monte-Carlo approach is that you have a
set of samples from the full parameter space. To answer any particular
question one can examine the points and compute results reliably,
taking full account of the shape of the posterior in $N$
dimensions. However for human consumption it is usual to summarize the
results as a set of parameter values and error bars. 

One way to do this is to use the samples for a principle component
analysis to identify the degeneracy directions, as we demonstrated in
Section~\ref{cmbonly}. By quoting constraints on a set of
orthogonalized parameters one retains most of the information in the
original distribution, as long as it is sufficiently Gaussian (or
Gaussian in the log, or some other function).
However ultimately one is usually interested in
the values of some fundamental parameters, and it is also useful to
find constraints on these alone. 

The simplest approach
is to compute the marginalized 1-dimensional
distributions for each parameter, essentially counting the number of
samples within binned ranges of parameter values. Note that this is
extremely hard to do using a brute-force numerical 
grid integration  
calculation as it
scales exponentially with the number of dimensions, but is quite
trivial from a set of Monte-Carlo samples. One can then quote the
value at the maximum or mean of the 1D distribution, along with extremal values of the
parameter
which contain a fraction $f$ of the samples, where $f$
defines the confidence limit. The extremal values could be chosen so
that there were the same number of outliers at both ends of the
distribution, or such that the value of the {\it marginalized}
probability is the same at each limit. This is a good way of
summarizing the current state of knowledge as long as you have
included everything you know, including using a believable prior over
parameter space.
%A comment(2): seems to me we have already made clear that
%marginalized isn't 'bad', just not what you want in some cases and
%that mean likelihoods gives different information.

However, frequently one wants to use the parameter estimates to assess
consistency with new data or theories, and the prior can be very hard to
define. 
For example, on putting in a top hat prior on the age and $h$, 
the {\it marginalized} prior probabilities are {\it not} flat,
even if all of the other priors are flat broad top hats.
This is because the marginalized
distribution includes the effect of the amount of parameter space
available at each point, which can depend quite strongly on the value
of the parameter. Likewise it is possible to have a region in
parameter space which fits the data rather well, but because the
region is small the 
marginalized probability of those
parameter values can be very low.
%A comment(2) : now maybe below
%Whilst certainly useful, results from the marginalized distribution should
%therefore be interpreted with care
%as they loose information about the shape of the full posterior and
%the goodness of fit.

When assessing consistency with new data (or theories), one really wants to know
whether the posterior for the new data intersects the $N$-dimensional
posterior for the current data in a region where both are likely. For example one could
define the region of parameter space enclosing a fraction $f$ of the
points with the highest likelihood as 
the $N$-dimensional confidence
region, and then see whether this region intersects with the corresponding
region for the new data. It is clearly sub-optimal to try to perform
this comparison using only 1D parameter values and limits, however if
one quotes the extremal values of each parameter contained in the $N$-dimensional confidence
region it is at least possible to assess whether the $N$-dimensional
regions \emph{might}
overlap. At least if the new data is outside these
limits it is a clear indication that there is an inconsistency,
whereas using the marginalized limits it shows no such thing (just
that if there is a consistent region it makes up a small fraction of
the original parameter space --- something one would hope for if the
new data is informative!) However it is of course easily possible for
the 1D likelihood limits to be consistent but the full $N$-dimensional
regions to be highly inconsistent.

%A moved
In order to be as informative as possible it can be useful to quote
both the marginalized and likelihood limits, though of course one
should study the full set of samples to make use of as much
information as possible. 
When there are strong degeneracies one can quote the constraints on
the well-determined orthogonalized parameters.

%A
\subsection*{Mean likelihoods}

Often it is useful to show the projected shape of the
distribution in one or two
dimensions. The marginalized distribution, proportional
to the number of samples at each point in the projected space, gives
the probability density in the reduced dimensions, ignoring the values
of the parameters in the marginalized dimensions,
and is therefore usually the quantity of interest. 
However
this looses all the information about the shape of the distribution in
the marginalized directions, in particular about the goodness of fit
and skewness with respect to marginalized parameters.
Useful complementary information is given by plotting the likelihood
of the best fit model at each point, for example see Ref.~\cite{Wang01}.
However it is not so easy to
compute this using a small set of Monte-Carlo samples ---
mean values within each bin can be obtained quite accurately from a
small number of samples, but getting a good value for the maximum in
each bin requires a much larger number. 
Instead we plot the mean likelihood of the samples at
each value of the parameter, which is easy to compute from the samples.
It shows how good a fit you could expect if you drew a random
sample from the marginalized distribution at each point in the subspace.

From a distribution $P(\vtheta)$ one can derive the (marginalized)
distribution of a derived parameter vector of interest $\vv=h(\vtheta)$ by 
\begin{equation}
P(\vv) = M(P,\vv) \equiv \int \d\vtheta \, P(\vtheta) \delta( h(\vtheta) - \vv).
\end{equation}
Assuming flat priors on $\vtheta$ the expected mean likelihood
of samples with $h(\vtheta)=\vv$ is
\begin{equation}
\la P(\vtheta: h(\vtheta)=\vv)\ra \equiv \frac{\int \d\vtheta\, P(\vtheta)^2 \delta(
h(\vtheta) - \vv)}
{\int \d\vtheta \,P(\vtheta) \delta(
h(\vtheta) - \vv)} = \frac{M(P^2,\vv)}{M(P,\vv)}.
\end{equation}
Frequently $h(\vtheta)$ is a projection operator into a subspace of
$\vtheta$ (for example  $h(\vtheta)=\theta_1$ for marginalization down
to the first parameter). If 
%A brackets above hopefully clear up ``this''
this 
is the case and $P(\vtheta)$ is a multivariate Gaussian distribution,
the marginalized distribution $M(P,\vv)$ is 
also a %S
Gaussian (readily proved
using Fourier transforms; the covariance is given by the projected
covariance matrix). 
Since the square of a Gaussian is a Gaussian 
%A
it follows that
$M(P^2,\vv) \propto
M(P,\vv)^2$, 
%S and hence 
%A then
%A I think you misread what I meant - added it follows that above
and hence
the mean likelihood is proportional to the
marginalized distribution $M(P,\vv)$. This also follows trivially if
$P$ is separable with respect to the subspace.
In the case of Gaussian or separable distributions the mean likelihood curve is therefore
proportional to the marginalized distribution and the two curves look the
same. 
Differences in the curves therefore indicate non-Gaussianity, 
for example when one of the marginalized parameters is skewing the
distribution in a particular direction (for example the effect of
massive neutrinos $f_\nu >0$ on the $\sigma_8$ curve in
Figure~\ref{inflation1D}; if $f_\nu$ was fixed at
it's maximum likelihood value the marginalized result for $\sigma_8$
would change significantly in the direction of the mean likelihood curve). 
The converse does not hold of
course, it is possibly to have a non-Gaussian distribution where both
curves are the same. If the priors on the parameters are not flat this
will also show up as differences in the curves even if the
likelihood distribution is Gaussian.

\subsection*{Effect of the prior}

In our analysis we chose a particular set of base parameters which were
assigned flat priors. This choice was fairly arbitrary, and there are
other possible choices. For example one might instead use $\Oml$ as a
base parameter and derive $h$ from the constraint
\begin{equation}
 h = \sqrt{\frac{\Omb h^2 + \Omc h^2}{1-\Oml-\OmK}}.
\end{equation}
In this case the prior on $h$ is given by 
\begin{equation}
P(h,\Omb h^2, \Omc h^2,\OmK) = P(\Oml,\Omb h^2, \Omc h^2,\OmK) \frac{\partial\Oml}{\partial h} = 2\frac{\Omm}{h}P(\Oml,\Omb h^2, \Omc h^2,\OmK),
\end{equation}
and so the prior on $h$ is proportional to $\Omm/h$ if the prior on $\Oml$
is flat. Using $h$ as a derived parameter therefore tends to give
results which favor lower $h$ values and higher $\Omm$ values.
Using importance sampling it is straightforward to adjust
results from one set of base parameters to another by weighting the
samples by the corresponding ratio of the priors\footnote{However the
tails of the distributions can change significantly, so it may be
necessary to generate the original chain at a higher temperature to
importance sample accurately. In this example we generated the chain
at a temperature of 1.3, so the samples were drawn from $P^{10/13}$
and then importance sampled.
}. 

\begin{figure}
\begin{center}
\BW{
\epsfig{figure=prior1_bw.ps,angle=0,width=7cm}
\epsfig{figure=prior2_bw.ps,angle=0,width=7cm}
}{
\epsfig{figure=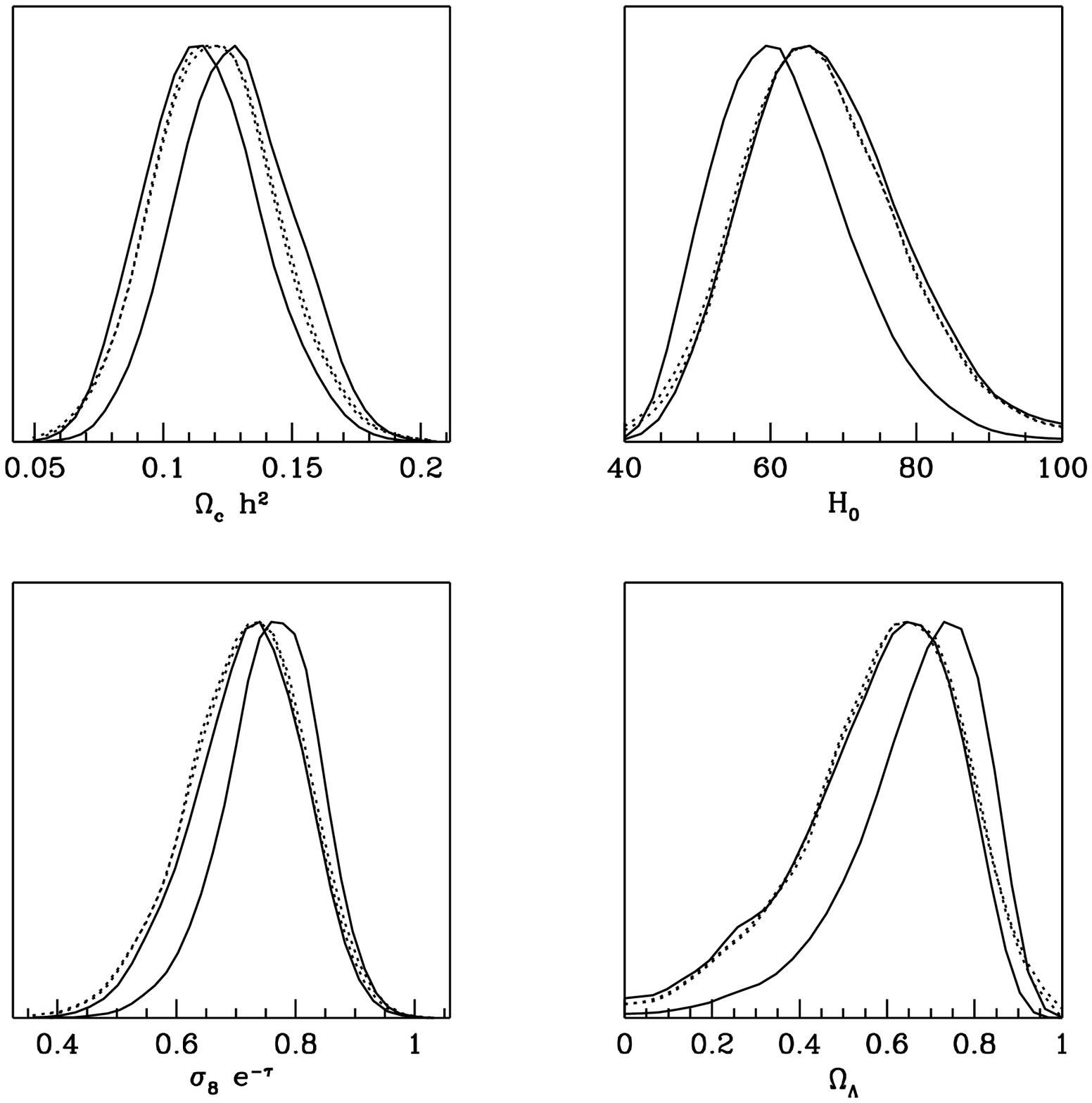,angle=0,width=7cm}
\epsfig{figure=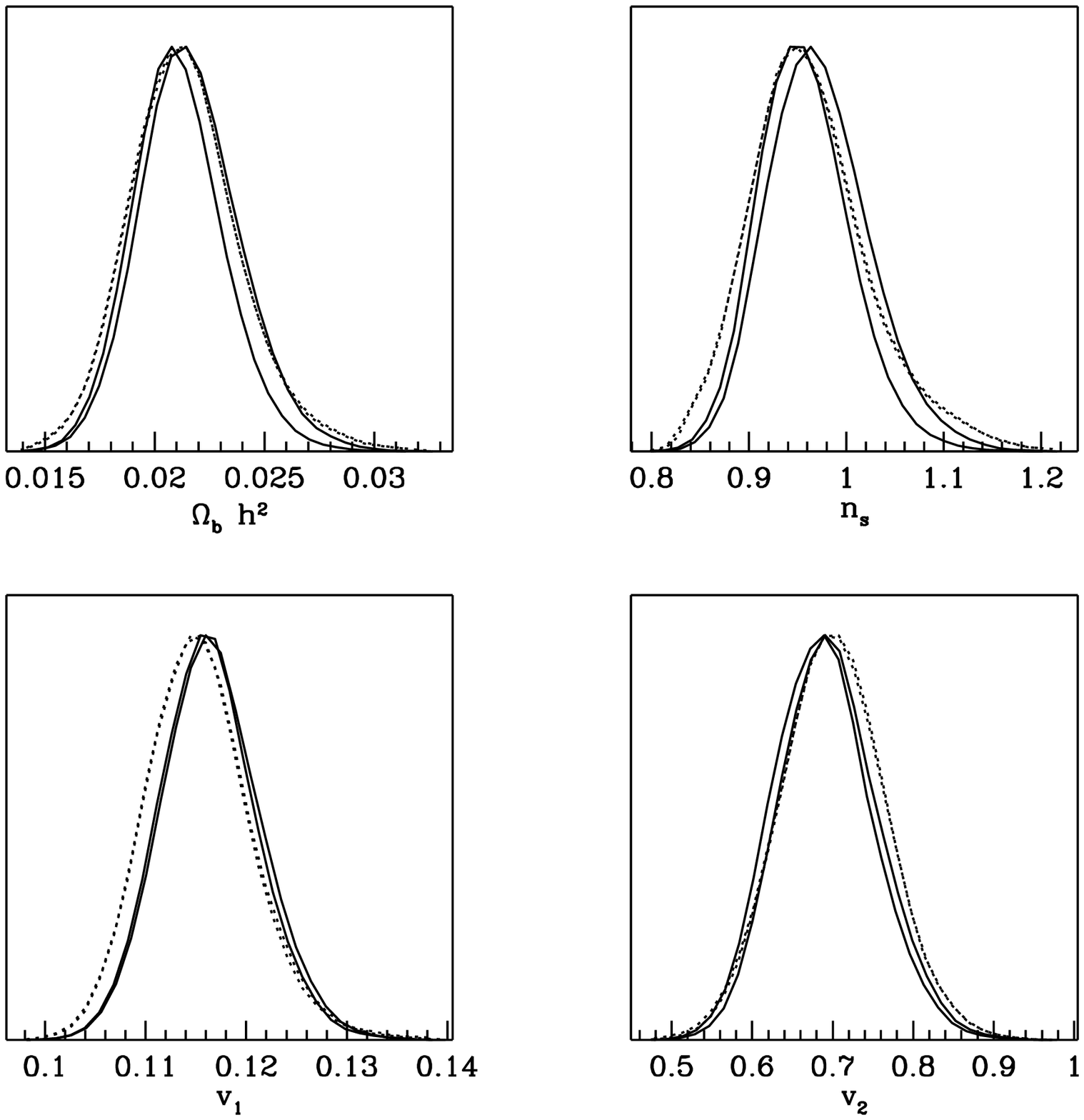,angle=0,width=7cm}
}
\caption{
Parameter constrains from the CMB alone with flat prior on $h$ ($\Oml$
derived, \BW{thick}{black} lines) and flat prior on $\Oml$ ($h$ derived, \BW{thin}{red}
lines). Dotted lines show the mean likelihood of the samples, solid
lines the estimated marginalized distribution. 
In most cases both sets of dotted lines are almost on top of one another.
}
\end{center}
\label{prioreff}
\end{figure}

For the results of
the parameter estimation to be meaningful it is essential that the
priors on the base set are well justified, or that the results are
independent of the choice. In Figure~\ref{prioreff} we show the effect
on the posterior constraints from the CMB data from 
the 6 parameter analysis using 
different base sets. The distributions shift by a fraction of
their width, though this can have quite a large effect on the
high-significance limits of weakly constrained parameters
(for example the $95\%$ confidence limit is $h < 0.89$ with $h$ a base
parameter, $h < 0.59$ with $\Oml$ a base parameter). 

For well constrained parameters the prior
effectively becomes flatter over the region of interest and the effect
is much less significant. As shown on the right of
Figure~\ref{prioreff} the posteriors of four parameters that are well constrained
by the CMB are almost independent of the choice of prior.

As shown in Figure~\ref{prioreff} plotting the mean
likelihood of the samples gives a 
clear indication of the direction in which results may be biased relative to a
different choice of prior. It is also clear that by choosing $h$ as
a base parameter we are getting more samples in the region of interest
for comparison with other data. In particular using $\Oml$ as a base
parameter gives a sharp cut-off at the higher values of $h$,
which are
 allowed
by the HST prior. 
One slight disadvantage of using $h$ rather than $\Oml$ is that 
the correlation of $h$ with some of the other base
parameters is significant, 
which may make the Monte-Carlo sampling less efficient.
However since we use the covariance matrix
to rotate to a set of orthogonalized parameters after one short initial run
this is not a major problem.

The efficiency of the MCMC implementation can be
improved by using a set of parameters for which the posterior is as symmetric as
possible~\cite{Kosowsky02}. It may therefore be a good idea to
transform to a better set of base parameters, 
for example one could transform to a set of orthogonalized parameters
derived from a principle component analysis using some less constraining data.
However when performing a non-linear transformation of parameters it is also
necessary to transform the flat priors on the parameters to obtain equivalent
results. If one assumes flat priors in the transformed parameters it
is wise to check that this does not induce a strong prior bias on the
cosmological parameters of interest.

\section{Goodness of fit}
\label{goodness}
To consider whether an enlarged parameter space is justified, one
ideally wants to compare the evidences $P(D)$ with the different
parameter sets. In some cases, for example when using hyperparameter
weights on experiments, it may be possible to define a prior on the
extra parameters in which case one can compute the evidence ratio
directly. The ratio does however depend quite strongly on the prior
put on the parameters, which in general it is not straightforward to
quantify. If one puts a broad prior on a parameter, but the
likelihood is much narrower, the probability of the data is
down-weighted because the likelihood takes up a much smaller region of
parameter space. One simple, but non-Bayesian, way to get round this
is to set the prior equal to the normalized posterior for computing the evidence, in which case one compare
the values of
\begin{equation}
\text{`$P(D)$'} = \frac{\int \d\vtheta P(D|\vtheta) P(\vtheta|D)}{\int \d\vtheta
P(\vtheta|D)} \approx \frac{1}{N}\sum_{n=1}^N P(D|\vtheta_n).
\end{equation}
This is just the expected probability of the data in the
posterior distribution, which can be estimated trivially from a set of
Monte-Carlo samples as the mean likelihood of the
samples. For Gaussian distributions this is the
 exponential mean of the chi-squareds under the posterior distribution, and
is thus a smeared-out version of the common practice of quoting the
chi-squared of the best fit. The smearing out
helps to down-weight extra parameters which have to be fine tuned to
obtain better fits. If the mean likelihood is bigger with the extra
parameters it suggests they are improving the fit to the data on average. Although
we know no way to use the value rigorously for hypothesis testing,
it seems nonetheless to be useful as a rule of thumb measure of
goodness of fit.

\section{Consistency of data sets}
\label{consistency}

It is important to assess whether the datasets 
being used are
consistent, or whether one or more is likely to be erroneous. This can
be done by introducing hyperparameter weights on the different
datasets~\cite{Lahav01,Hobson02} when performing the analysis. If a dataset is
inconsistent, its posterior hyperparameter 
will have a low value, and
the dataset then only contributes weakly to the posterior probability
of the parameters. 
In the
case that the likelihoods are of Gaussian form it is a simple matter to
marginalize over the hyperparameters analytically given a simple prior. To assess whether
the introduction of hyperparameters is justified (i.e. whether the data
are inconsistent with respect to the model), one can compare the
probability of obtaining the data in the two hypotheses: $H_0$, no
hyperparameters are needed; $H_1$, hyperparameters are needed because one or more
datasets are
inconsistent. Using a maximum entropy prior assuming that
on average the hyperparameter weights are unity, Ref.~\cite{Hobson02} gives
\begin{equation}
\frac{P(D|\vtheta,H_1)}{P(D|\vtheta,H_0)} =
\prod_k \frac{2^{n_k/2+1}\Gamma(n_k/2+1)}{(\chi^2_k +2) e^{-\chi_k^2/2}},
\end{equation}
where $k$ labels the datasets, each 
containing
$n_k$ points. Given a set of independent samples from $H_0$ it
is straightforward to compute an estimate of the evidence ratio using Eq.~\eqref{evid_ratio}.
If the datasets are inconsistent the importance sampling estimate
would be very inaccurate as the probability distributions would be
significantly different. However this should be clear when one
computes the estimate since the probability ratios will vary wildly. If
one suspects that one of the datasets is inconsistent it would
be better to start with sampling from $H_1$, and confirm that
the evidence ratio supports using the hyperparameters.

An even simpler way of assessing consistency of the datasets 
might be to compare the mean likelihood of the samples in the region of
overlap of the posterior distributions to the overall mean likelihood
under the original posterior. 
If the mean likelihood of the samples in the region of overlap is much less than the 
original mean,
it is an indication than the regions of high likelihood under each
dataset do not overlap well in $N$-dimensions, and hence there may be
an inconsistency. In practice the samples in the region of overlap can
be found by importance sampling additional datasets.
The mean likelihoods 
should always be computed with respect to the 
same, original, dataset
(or group of datasets).
However importance sampling may fail to identify inconsistencies in
particular cases when the distributions have multiple maxima.

\section{Analytic marginalization}
\label{marge}
Frequently one has data in which there is an unknown calibration
uncertainty, or an unknown normalization. These
parameters can be marginalized over analytically
following~\cite{Bridle01} as long as the likelihoods are Gaussian, and
the prior on the amplitude parameter is Gaussian or flat. Typically one has
a marginalization of the form
\begin{equation}
L \propto \int \d\alpha P(\alpha) \exp[- (\alpha \vv -\vd)^T \mN^{-1} (\alpha \vv -\vd) /2]
\end{equation}
where $\vv$ and $\vd$ are vectors, $\mN$ is the noise covariance matrix, and
$P(\alpha)$ is the prior. For example for the supernovae data $\vv$
is assumed to be a vector of equal constants giving the intrinsic magnitudes
of the supernovae, and $\vd$ is a vector of the theoretical minus the
observed effective magnitudes. If the prior $P(\alpha) = \text{const}$ it
clearly cannot be normalized, however the marginalization is trivial
giving
\begin{equation}
- 2\ln L =  \vd^T\left( \mN^{-1} - \frac{\mN^{-1}  \vv \vv^T \mN^{-1}}{\vv^T
\mN^{-1} \vv} \right ) \vd + \ln(\vv^T \mN^{-1} \vv)  + \text{const}.
\end{equation}
In the case that $\vv$ is a constant (independent of the data and
parameters), one has $L\propto e^{-\chi^2_{\text{eff}}/2}$ where
\begin{equation}
\chi^2_{\text{eff}} = \vd^T\left( \mN^{-1} - \frac{\mN^{-1}  \vv \vv^T \mN^{-1}}{\vv^T
\mN^{-1} \vv} \right) \vd = \chi^2_{\text{best-fit}}.
\end{equation}
This is exactly the same as the best-fit one obtains by
minimizing the likelihood w.r.t. $\alpha$, and so in this case the
maximization technique of Ref.~\cite{Wang01} is exactly equivalent to
full marginalization. For example, in the case of the
supernovae data, marginalization with a flat prior over the magnitudes
is equivalent to using the best-fit magnitude. In general this is not true as the
logarithmic dependence $\ln(\vv^T \mN^{-1} \vv)$ can depend on the
parameters. For example with the 2dF data $\vv$ would be the predicted
matter power spectrum values, and $\alpha$ would be the unknown
amplitude relative to the galaxy power spectrum at $z=0.17$.

The marginalized result is only `correct' if the
assumed flat prior is correct; it is an advantage of the maximization
technique that the result does not depend on the prior.

%%%%%%%%%%%%%%%%%%%%%%%%%%%%%%%%%%%%%%%%%%%%%%%%%%%%%%%%%%%%%%%%%%
%                        References                              %
%%%%%%%%%%%%%%%%%%%%%%%%%%%%%%%%%%%%%%%%%%%%%%%%%%%%%%%%%%%%%%%%%%

%\bibliography{../antony}

\end{document}